\documentclass[prb,reprint,superscriptaddress,amsmath,amssymb,aps]{revtex4-2}

\AtBeginDocument{%
  \geometry{top=1.0in,bottom=1.0in,left=0.8in,right=0.8in}%
}

\usepackage{graphicx}
\usepackage[utf8]{inputenc}
\usepackage{feynmp-auto}
\usepackage{physics}
\usepackage{xcolor}
\usepackage[a4paper, margin=1in]{geometry}
\usepackage{verbatim}
\usepackage{mathtools}
\usepackage{amsmath}
\usepackage{amssymb}
\usepackage{pifont}
\usepackage[percent]{overpic}
\usepackage{enumitem}
\usepackage{hyperref}
\hypersetup{colorlinks=true,urlcolor=blue,citecolor=blue,linkcolor=blue,breaklinks=true}
\usepackage[noabbrev]{cleveref}
\usepackage{dcolumn}

\begin{document}

\newcommand{\bfk}{\mathbf{k}}
\newcommand{\bfq}{\mathbf{q}}
\newcommand{\X}{\mathrm{X}}
\newcommand{\SC}{\mathrm{SC}}
\newcommand{\tSC}{\mathrm{tSC}}
\newcommand{\sSC}{\mathrm{sSC}}
\newcommand{\dSC}{\mathrm{dSC}}
\newcommand{\dM}{\mathrm{dM}}
\newcommand{\dD}{\mathrm{dD}}
\newcommand{\D}{\mathrm{D}}
\newcommand{\M}{\mathrm{M}}
\newcommand{\pp}{\mathrm{pp}}
\newcommand{\ph}{\mathrm{ph}}
\newcommand{\xph}{\mathrm{xph}}

\newcommand{\dwave}{\text{\scriptsize d-wave}}

\newcommand{\guess}{\text{\scriptsize i=0}}

\newcommand{\sigin}{\sigma}
\newcommand{\sigout}{{\sigma'}}
\newcommand{\sigintilde}{\tilde{\sigma}}
\newcommand{\sigouttilde}{{\tilde{\sigma}'}}

\newcommand{\calQ}{\mathcal{Q}}

\title{Diagrammatic bosonization, aspects of criticality, and the Hohenberg-Mermin-Wagner Theorem in parquet approaches}
\author{Aiman Al-Eryani}
\affiliation{Theoretical Physics III, Ruhr-University Bochum, 44801 Bochum, Germany}

\begin{abstract}
   The parquet equations present a cornerstone of some of the most important diagrammatic many-body approximations and methods currently on the market for strongly correlated materials: from non-local extensions of the dynamical mean-field theory to the functional renormalization group. The recently introduced single-boson exchange decomposition of the vertex presents an alternative set of equivalent equations in terms of screened interactions, Hedin vertices, and rest functions. This formulation has garnered much attention for several reasons: opening the door to new approximations, for avoiding vertex divergences associated with local moment formation plaguing the traditional parquet decomposition, and for its interpretative advantage in its built-in diagrammatic identification of bosons without resorting to Hubbard-Stratonovich transformations. In this work, we show how the fermionic diagrams of the particle-particle and particle-hole polarizations in the SBE formalism can be mapped to diagrammatics of a bosonic self-energy of two respective bosonic theories with pure bosonic constituents, solidifying the identification of the screened interaction with a bosonic propagator. Resorting to a spin-diagonalized basis for the bosonic fields and neglecting the coupling between singlet and triplet components is shown to recover the trace log theory known from Hubbard-Stratonovich transformations. Armed with this concrete mapping, we revisit a conjecture claiming that universal aspects of the parquet approximation coincide with those of a self-consistent large-$N$ approximation for a bosonic $O(N)$ model. We comment on the role of the self-energy and crossing symmetry in enforcing the Hohenberg-Mermin-Wagner theorem in parquet-related approaches. 
\end{abstract}

\maketitle

\section{Introduction}

The last decade has marked a flurry of experimental discoveries of new families of quantum materials featuring strongly correlated electrons, such as materials with perovskite structure whether cuprates~\cite{cupratesfirst_Bednorz1986,cupratessecond_PhysRevLett.58.908} or ruthenates~\cite{perovskiteruthenates_Maeno1994}, iron-based superconductors~\cite{ironsc_Kamihara2008}, nickelates~\cite{infinitelayernickelates_Li2019} and more recently, strongly correlated layered van der Waals materials~\cite{vanderwaals_Cao2018_super,vanderwaals_Cao2018_insulator,vanderwaals_Park2021_multilayer,vanderwaals_Cao2021_MATTG,vanderwaals_Xie2022_promotion,vanderwaals_Xia2023_tWSe2}. These materials exhibit a rich landscape of competing phases -- magnetism, superconductivity, and charge ordering -- driven by strong electronic correlations. Developing a robust theoretical understanding of these phenomena remains a major challenge due to the intricate interplay of fluctuations. Among the most promising theoretical approaches to tackle this problem are diagrammatic methods, which offer a systematic framework for going beyond mean-field theories and capturing the complex correlations in these systems. A particularly powerful toolkit in this context is the parquet formalism, which re-sums a class of Feynman diagrams capturing the feedback between different fluctuation channels -- notably, particle-hole and particle-particle processes -- while respecting fundamental symmetries such as crossing symmetry -- a consequence of the Pauli exclusion principle. The parquet approximation (PA), developed in the context of nuclear and condensed matter physics \cite{diatlov_sudakov_TerMartirosian_parquet_foundation_1957,dominics_martin_parquet_foundation_1964}, has seen a resurgence thanks to recent developments in computational techniques and vertex parametrization in momenta \cite{Lichtenstein_2017,Huseman_salmhoefer_omega_flow_and_effeicient_parametrization} and frequencies \cite{highfreq_nils_asymptotics,Karrasch_2008}. It has since been extended and combined with other many-body methods, such as the dynamical vertex approximation (D$\Gamma$A) \cite{DGammaA2007,abinitio_dgamma_a} which extends dynamical mean field theory\cite{DMFT_original1_MetznerVollhardt1989,DMFT_original2_GeorgesKotliarKrauthRozenberg1996} (DMFT), and has also been linked to the functional renormalization group (fRG) \cite{Metzner_2012_frg_review,Platt_frg_multiorbital_2013,kopietz2010introduction_book} in the so-called multiloop extension \cite{multiloop_from_parquet_Kugler_2018,Gievers_2022,kiesePhysRevResearch.4.023185} that can be seen also in a systematic convergence in including certain higher loop corrections to the flow equations.

A recent development is the single-boson exchange (SBE) decomposition of the vertex~\cite{Krien_SBE_original,Krien_2019}, which offers a physically transparent parametrization based on the diagrammatic concept of bare-interaction reducibility. The SBE decomposition has attracted significant attention because it avoids the problem of vertex divergences associated to local moment formation -- a problem that has hampered methods relying on the traditional parquet decomposition of the vertex~\cite{vertexdiv1_PhysRevLett.110.246405,vertexdiv2_PhysRevB.94.235108,vertexdiv3_PhysRevLett.119.056402,vertexdiv4_PhysRevB.97.245136,vertexdiv5_PhysRevB.98.235107,vertexdiv6_PhysRevB.101.155148,vertexdiv7_Chalupa_2021} -- as shown in~\cite{smokinggun_10.21468/SciPostPhys.16.2.054}. Moreover, the SBE decomposition naturally identifies Hedin-like screened interactions as bosonic propagators within fermionic diagrammatics, further reinforcing its advantageous physical transparency. 

It has since been noticed that functional renormalization group equations of a multi-channel Hubbard-Stratonovich transformation (HST) treatment of fermionic theories~\cite{Denz_PhysRevB.101.155115} coincide with the flow equations of these screened interactions of the SBE bosonic propagators~\cite{SBE_strong_Bonetti_2022,fraboulet2023singlebosonexchangefunctionalrenormalization}. This hints that the SBE polarizations may coincide with the bosonic self-energies of the HST theory after fermions have been integrated out. In this work, we solidify the bosonic interpretation of the SBE screened interactions and polarizations, and show this connection to the HST concretely from a diagrammatic point of view. In particular, we give a procedure of diagrammatic bosonization, showing how fermionic diagrams of a screened interaction and polarization diagrams of a given channel $r$ can be consistently mapped to ones of a pure bosonic theory $S^r[\varphi^r]$. To show the utility of such a mapping, we revisit the problem of finite temperature critical properties of parquet approaches. In particular, an important physical constraint in exact solutions to a many body model of fermions is the Hohenberg-Mermin-Wagner (HMW) theorem \cite{hohenberg_mermin_wagner_theorem,mermin_wagner_thm_original}, proven for lattice fermion systems in Ref.~\cite{Koma_tasaki_mermin_wagner_thm}, which forbids long-range order in low-dimensional systems due to strong fluctuations in low dimensions which destroy the order. Although realistic considerations can easily hinder its applicability~\cite{Palle_2021}, the fulfillment of this constraint constitutes an important benchmark of many-body approximations. In approximations within the parquet formalism, assessing the degree to which such fluctuations are properly captured is subtle, as the theory explicitly incorporates two- and four-point fermionic interactions, corresponding to one- and two-point bosonic functions, but offers limited transparency regarding the presence of higher-order bosonic interactions that can influence critical behavior.

Previous theoretical work by Bickers and Scalapino \cite{parquet_critical_bickers_scalapino_1992} addressed this question for the PA by conjecturing that at criticality, it is connected to a bosonic approximation of an $O(N)$ model for the critical fluctuations called the self-consistent screening approximation (SCSA) \cite{bray_scsa_original_1974}. They presented a heuristic argument for this connection, and argued that as a consequence PA solutions respects the HMW theorem. A concrete mapping of the fermionic parquet to bosonic diagrammatics necessary for this argument however remained lacking. We are able to finally address this gap.  By appealing to power counting arguments near criticality, we demonstrate that PA solutions near criticality are in the same universality class as bosonic SCSA solutions. We show the role of the self-energy and crossing symmetry in enforcing the HMW theorem. We further comment on other related approximations.   

\section{Method}

\subsection{The Single Boson Exchange Equations}
\label{sec:sbe_intro}
\begin{figure}
    \centering
    \includegraphics[width=1.0\linewidth]{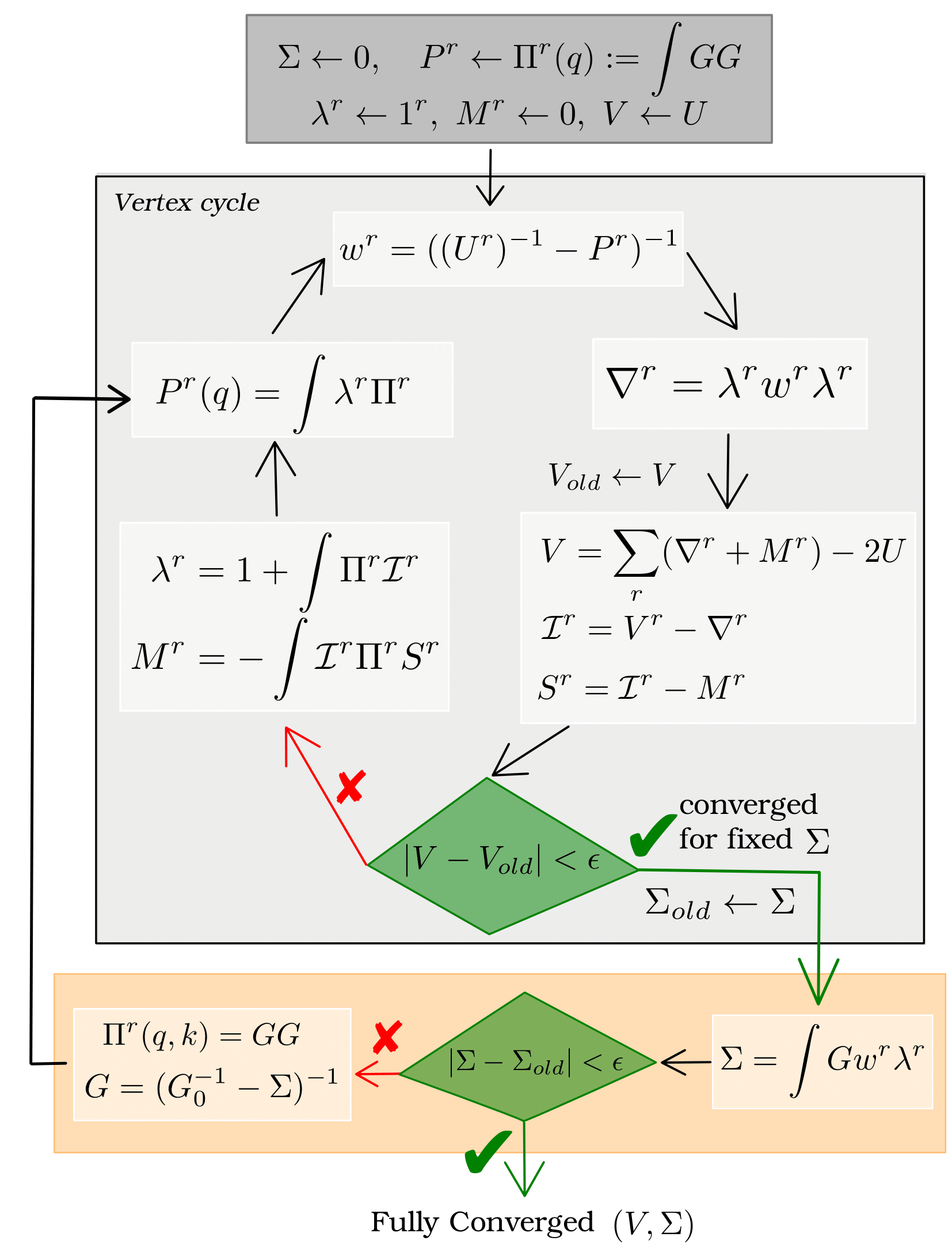}
    \caption{A possible solution strategy for the Parquet equations in the SBE formalism where the vertex cycle is decoupled from the self-energy update~\cite{plain_and_simple_parquet_Krien_2022}.}
    \label{fig:solution_strategy}
\end{figure}

Let $\psi_{\sigma k}, \bar{\psi}_{\sigma k}$ be Grassmann fields corresponding to electron creation and annihilation operators. Here $\sigma \in \{\uparrow,\downarrow\}$ is a spin argument while $k = (i\nu, \bfk)$ is a compound frequency-momentum index containing a fermionic Matsubara frequency $i\nu$ and a momentum $\bfk$. We denote the integration by these indices as $\int_{\sigma k} = \sum_{\sigma}\sum_{i\nu}\int_{\bfk}$, where the normalization factor by the volume of one over the BZ/number of sites $N$ on the reciprocal lattice is absorbed in the notation of the integral sign, and a normalization factor of $1/\beta$ is absorbed in the Matsubara summation. The general fermion Euclidean action
\begin{align}
S[\psi, \bar{\psi}] =& -\int\limits_{\overset{k_1k_2}{\sigma_1\sigma_2}} \bar{\psi}_{\sigma_1k_1} G^{-1}_0(k_1, k_2)_{\sigma_1\sigma_2} \psi_{\sigma_2k_2} \nonumber \\ 
&+ \frac{1}{4} \hspace{-3mm}\int\limits_{\overset{k_1k_2k_3k_4}{\sigma_1\sigma_2\sigma_3\sigma_4}}\hspace{-4mm} V_0 \psi_{\sigma_1k_1}\bar{\psi}_{\sigma_2k_2}\psi_{\sigma_3k_3}\bar{\psi}_{\sigma_4k_4},
\end{align}
describes the Hubbard model for the bare Green's function $G_0(k_1, k_2)_{\sigma_1\sigma_2} := G_0(k_1)_{\sigma_1\sigma_2}\delta_{k_1, k_2} := G_0(k_1)\delta_{k_1, k_2}\delta_{\sigma_1\sigma_2}$ defined by
\begin{align}
    &G_0(k) = \frac{1}{i\nu - \epsilon(\bfk)}, \\ & \epsilon(\bfk) = 2t(\cos(k_x) + \cos(k_y)) - \mu
\end{align}
and $V_0 = V_0(k_1, k_2, k_3, k_4)_{\sigma_1\sigma_2\sigma_3\sigma_4}$ the bare interaction defined by
\begin{align}
V_0 = U \delta_{\bfk_1 + \bfk_3, \bfk_2 +\bfk4}(\delta_{\sigma_1,\sigma_2} \delta_{\sigma_3,\sigma_4} - \delta_{\sigma_1, \sigma_4} \delta_{\sigma_3, \sigma_2}), \label{eq:bare_vertex_spin_invt}
\end{align}
with the first delta function implementing momentum conservation by translation symmetry and the second implementing the SU(2)-spin symmetry of the Hubbard model. 
Let $A(k_1, k_2, k_3, k_4)_{\sigma_1\sigma_2\sigma_3\sigma_4}$ be a two-fermion object. Here $j=1,3$ represent ingoing legs, and $j=2,4$ represent outgoing legs. Assuming thermodynamic equilibrium and translation invariance of the theory, meaning $A = A\delta_{k_1+k_3, k_2+k_4}$, then $A$ generally depends on only three frequency-momentum variables, and it is customary to define three channel parameterizations
\begin{align}
A^\pp_{kk'}(q)_{\sigin \sigintilde ;\sigout \sigouttilde} &= A(k, k', q-k, q-k')_{\sigin\sigout\sigintilde\sigouttilde}\\
A^\ph_{kk'}(q)_{\sigin \sigintilde ;\sigout \sigouttilde} &= A(k, k+q, k'+q, k')_{\sigin\sigintilde  \sigouttilde \sigout}\\
A^\xph_{kk'}(q)_{\sigin \sigintilde ;\sigout \sigouttilde} &= A(k, k', k'+q, k+q)_{\sigin\sigout\sigouttilde\sigintilde},
\end{align}
where $q = (i\Omega, \bfq)$ is a transfer bosonic-frequency momentum variable with $i\Omega$ a bosonic Matsubara frequency and $k = (i\nu, \bfk)$, $k' = (i\nu', \bfk')$ are secondary fermionic ones. Then, we let the product of $A$ with another two-fermion object $B$ in the same diagrammatic channel $r \in \{\pp, \ph, \xph\}$ represent matrix multiplication in the spin indices
\begin{align}
(A^r   B^r)_{\sigin \sigintilde ;\sigout \sigouttilde} = \sum_{s \tilde{s}} A^r_{\sigin \sigintilde ;s \tilde{s}} B^r_{s \tilde{s} ;\sigout \sigouttilde}.
\end{align}
Given a fermionic propagator of a translationally invariant system $G(k)_{\sigin \sigout}$, we define the particle-particle and particle-hole bubbles as~\cite{Gievers_2022}
\begin{align}
\Pi^\pp_{kk'}(q)_{\sigin\sigintilde;\sigout\sigouttilde} &= \frac{1}{2}\delta_{k,k'} G(k)_{\sigin\sigouttilde}G(q-k)_{\sigintilde\sigout}\\
\Pi^\ph_{kk'}(q)_{\sigin\sigintilde;\sigout\sigouttilde} &= -\delta_{k,k'} G(k)_{\sigin\sigouttilde}G(q+k)_{\sigout \sigintilde}\\
\Pi^\xph_{kk'}(q)_{\sigin\sigintilde;\sigout\sigouttilde} &= \delta_{k,k'} G(k)_{\sigin\sigout}G(q+k)_{\sigouttilde \sigintilde},
\end{align}
with the full Green's function determined from the bare Green's function $G_0$ and the electron self-energy $\Sigma(k)_{\sigma\sigma'}$ using the Dyson equation 
\begin{align}
G(k)_{\sigma\sigma'} = [G_0(k) - \Sigma(k)]^{-1}_{\sigma\sigma'}, \label{eq:dyson_equation}
\end{align}
with the inverse taken in spin space. Having SU(2)-spin symmetry implies $G$, $G_0$ and $\Sigma$ are proportional to the identity in spin space. 

To ensure identical equations later for any $r$, we need to additionally define channel specific two-particle identities in spin space, which are trivial in their respective channel parametrization
\begin{align}
\mathbf{1}^\ph_{\sigin\sigintilde;\sigout\sigouttilde} = \mathbf{1}^\xph_{\sigin\sigintilde;\sigout\sigouttilde} =  \mathbf{1}^{pp}_{\sigin\sigintilde;\sigout\sigouttilde} = \delta_{\sigin\sigout}\delta_{\sigintilde\sigouttilde}. 
\end{align}

A set of self-consistent equations for the SBE splitting of the vertex have been derived and are formally equivalent to the parquet equations ~\cite{Krien2021b_tiling_with_triangles,plain_and_simple_parquet_Krien_2022}. We list these equations while describing a scheme for how they can be solved self-consistently~\cite{Krien_2019,Krien2021b_tiling_with_triangles,plain_and_simple_parquet_Krien_2022}. For a graphical representation, see Fig.~\ref{fig:solution_strategy}. First, an initial guess for the self-energy $\Sigma$, polarizations $P^r(q)$, Hedin vertices $\lambda^r_k(q)$, rest functions $M^r_{kk'}(q)$, as well as the full vertex $V$ is taken, where the spin indices have been suppressed. The initial values are chosen to be
\begin{align}
\Sigma_{\guess} &= 0,\  P^r_{\guess} := \int_{kk'}\Pi_{kk'}^r(q), \nonumber \\ \lambda^r_{\guess} &= \mathbf{1}^r,\  M^r_{\guess} = 0,\ V^r_{\guess} = V_{\text{2PI}}^r,
\end{align}
where $V_{\text{2PI}}^r$ is an expression for the two-particle irreducible part of the vertex which is needed to be provided as input -- typically in an approximate fashion.

As a first step, the bosonic propagator is calculated using the Dyson equation
\begin{subequations}
\label{eq:vertex_system_of_eqns}
\begin{align}
     w^r(q) &= U^r - U^r   P^r(q)   w^r(q)  \nonumber \\
     &= (\mathbf{1}^r+U^rP^r(q))^{-1}U^r,\label{eq:screened_interaction}
\end{align}
either by iteration of the first equality finitely many times, or by a single evaluation of the second equality. It is suggested to use the second equality since the first is unstable due to the possibility of a small denominator in the process of convergence~\cite{kiese2024embeddedmultibosonexchangestep}. We found fastest convergence to occur when the second equation is used for sub-leading channels and iteration of the first equation for the leading channel(s), if known apriori. Next, the SBE contribution to the vertex is calculated using
\begin{align}
\nabla^r_{kk'}(q) = \lambda^r_k(q)   w^r(q)    \lambda_{k'}(q).
\end{align}
Then the old vertex, which we denote by $V_{old}$ is stored before a new vertex can be updated using
\begin{align}
     V^r &= \sum_{r'} (\nabla^{r'} + M^{r'} - U^{r'}) + V_{\text{2PI}}^r \label{eq:SBE_decomposition}\\
     \mathcal{I}^r &= V^r - \nabla^r \label{eq:U_irreducible_vertex}
\end{align}
where $\mathcal{I}^r$ is the U-irreducible vertex in the channel $r$. The Hedin vertices are then updated using~\cite{Krien_2019}
\begin{align}
     \lambda^r_k(q) &= \mathbf{1}^r + \int_{k'k''} \Pi^r_{k'k''}(q) 
       \mathcal{I}^r_{k''k}(q) \label{eq:hedin_vertex}
\end{align}
and the rest functions using
\begin{align}
M^r_{kk'}(q) &= -\int_{k''k'''} \mathcal{I}^r_{kk''}(q)   \Pi^r_{k''k'''}(q)   S_{k'''k'}^r(q) \nonumber \\ \qq{with} &S^r = \mathcal{I}^r - M^r. \label{eq:restfunction_equation}
\end{align}
With the Hedin vertices, one can update the polarizations using
\begin{align}
     P^r(q) &= \int_{kk'} \lambda^r_k(q)  \Pi^r_{kk'}(q), \label{eq:polarization}
\end{align}
which can be fed back into Eq.~\eqref{eq:screened_interaction} to calculate the new screened interactions. Although we have not updated the self-energy, this system given in Eqs.~\eqref{eq:vertex_system_of_eqns} represent a closed system of equations for the vertex $V$. This cycle of calculations, represented in the light grey box in Fig.~\ref{fig:solution_strategy}, can be repeated until a convergence criterion $||V - V_{old}||/||V|| < \varepsilon_V$, with $V_{old}$ the vertex calculated in a previous iteration, and $\varepsilon_V$ a convergence threshold, is satisfied. Then one says to have converged at the level of the vertex. Note, however, that since the self-energy is left without an update, such a solution would not yet constitute a full parquet solution. 

\end{subequations}
For a fully consistent calculation, we need to further update the self-energy using the Schwinger-Dyson equation (SDE)~\cite{Patricolo_2025}
\begin{align}
\Sigma(k)_{\sigma_1\sigma_2} = \int_{q\sigma\sigma'} \hspace{-0.3cm} (w^r(q) \lambda^r_{q-k}(q))_{\sigma_1 \sigma;\sigma' \sigma_2} G(q - k)_{\sigma\sigma'}, \label{eq:SDE_SBE}
\end{align}
where $r \in \{\pp, \ph, \xph\}$ can be chosen freely (e.g. to minimize the effect of further truncations).
The self-energy is stored before the update and denoted by $\Sigma_{old}$. If the new self-energy changes from the old self-energy by a large extent; i.e. one does not have $||\Sigma - \Sigma_{old}||/||\Sigma|| < \varepsilon_\Sigma$, then the update in the self-energy will result in a sizable change of the Green's function $G$, and therefore a sizable change of the bubbles $\Pi^r$. Since the bubbles show up everywhere in the equations constituting the vertex cycle Eqs.~\eqref{eq:vertex_system_of_eqns}, one has in general that the vertex is no longer converged, and the vertex cycle must be converged again. This is repeated until the solution $(V, \Sigma)$ is converged at the level of both the vertex and the self-energy. 

The scheme described here where the vertex cycle is converged initially before the self-energy updated is not the only way, neither the best way in practice, to converge the parquet equations. Indeed, one can include the self-energy update after every vertex cycle, and test for convergence in the vertex and the self-energy at the same time. Another scheme is to replace $G_0$ or $V_0$ by a one-parameter family $G_0^\Lambda$ or $V_0^\Lambda$ and take a derivative of the flow equations with respect to $\Lambda$, to obtain a set of ODEs for $w^r$, $\lambda^r$, $M^r$ and $\Sigma$ which can be solved instead, making the connection to functional RG equations \cite{Gievers_2022,kfraboulet_multiloop}. The distinguishing feature of the scheme we have chosen for this work: that in the vertex cycle the self-energy is kept constant, will be important to our discussion.

At this point, it should be mentioned that the equations~\eqref{eq:vertex_system_of_eqns} can also be represented in the so-called physical channels, in which an object of the $SU(2)$-symmetric system in the particle-hole or a particle-particle parametrization can be be spin-diagonalized to yield singlet and triplet components, thereby simplifying the spin structure of the object. For objects in the $r = \ph$ channel, the singlet component is identified with the charge or density channel $\D$, and the triplet component is identified with the spin or magnetic channel $\M$. Similarly, for objects in the $r=\pp$, the singlet (triplet) component is identified with the singlet  (triplet) pairing or superconducting channel $\mathrm{sSC}$ ($\mathrm{tSC}$). See Refs.~\cite{Patricolo_2025,Bickers2004bookchapter} for more details. Given a quantity $A^r$, we denote by $A^\X$ the physical channel and component, where $\X \in \{\D, \M, \mathrm{sSC}, \mathrm{tSC}\}$.

Solving the parquet equations present various difficulties: the first is to provide a good input $V_{\text{2PI}}$. In principle, it would be expected that with the exact $V_{\text{2PI}}$ that iteration of these equations should converge to the exact solution. In practice, this is typically obstructed by the existence of unphysical solutions, numerical effort and the numerical stability of the equations, although efforts to address these problems are subject of ongoing research~\cite{vertexdiv8_essl2025stayphysicalbranchselfconsistent,kiese2024embeddedmultibosonexchangestep,lihm2025finitedifferenceparquetmethodenhanced}. Therefore, various approximations have been utilized. We note here a common few in descending levels of complexity.
\begin{enumerate}[label={},leftmargin=0pt] 
    \item \textit{The parquet approximation~\cite{plain_and_simple_parquet_Krien_2022,PA4x4_PhysRevE.80.046706,diatlov_sudakov_TerMartirosian_parquet_foundation_1957,dominics_martin_parquet_foundation_1964} (PA)}. In the PA approximation, we approximate the input by the bare interaction $V_{\text{2PI}} \approx U$. The diagrammatic content of the vertex in this approximation thus excludes all non-trivial 2-particle irreducible vertex diagrams.   
    \item \textit{The SBE(b) approximation~\cite{Krien_2019}}~\footnote{The (b), is to distinguish this approximation from the SBE(a) approximation used in the single-boson exchange formulation of the related functional RG equations~\cite{fraboulet2023singlebosonexchangefunctionalrenormalization}, where the \emph{flow} of $M^r$ is neglected. These two approximations are distinct, and in general, it was found that the SBE(a) approximation outperforms the SBE(b). For the rest of this article, we drop the ``(b)'' and refer to it simply as the SBE approximation.}. On top of the PA approximation, the update of $M^r$ in Eq.~\eqref{eq:restfunction_equation} is skipped entirely, thereby dropping all non-trivial $U$-irreducible diagrams. This greatly reduces the numerical effort. This approximation can be interpreted to correspond to neglecting multiboson diagrams in the vertex. 
    \item \textit{The FLuctuation EXchange~\cite{Bickers1989_FLEX1,Bickers1989_FLEX2} (FLEX) approximation.} On top of the SBE(b) approximation, the update of the Hedin vertices Eq.~\eqref{eq:hedin_vertex} is skipped. The full vertex then becomes a sum of bosonic propagators, and is called the fluctuation exchange propagator. Feedback between the channels then enter only through the self-energy $\Sigma$. 
    
    \item \textit{The GW~\cite{GWoriginal_PhysRev.139.A796} approximation.} Like the FLEX, but one updates the screened interaction of a single preferred physical channel $w^\X$ only, with the self-energy updated using the respective channel equation, hence inter-channel feedback is absent entirely. The self-energy equation then takes the form  $\Sigma = \int_q G w^\X$; hence the name of the approximation. The diagrammatic content of the vertex in this approximation is a ladder summation in the preferred channel $\X$ where the electron Green's function lines appearing in these ladders are self-consistently dressed with the aforementioned self-energy. 

    \item \textit{Others.}  Closely related to the PA is the D$\Gamma$A~\cite{DGammaA2007,Eckhardt_2020,lihm2025finitedifferenceparquetmethodenhanced} where $V_{\text{2PI}} \approx V^{\text{DMFT}}_{\text{2PI}}$, where $V^{\text{DMFT}}_{\text{2PI}}$ is the local 2PI vertex calculated within DMFT, and the Embedded multi-boson exchange~\cite{kiese2024embeddedmultibosonexchangestep} (eMBEX) which forfeits the step of calculating the rest functions Eq.~\eqref{eq:restfunction_equation} and approximates $\Lambda_{\text{Uirr}} = V_{\text{2PI}} + \sum_r M^r$ -- the bare interaction-irreducible part of the vertex -- with a solution of the dynamical cluster approximation~\cite{DCA_Hettler_2000} (DCA) in a self-consistent fashion.  
\end{enumerate}

\section{Fermionic to Bosonic Mapping}

\subsection{Exact Bosonic Theory for the Screened Interaction} \label{sec:exact_bosonic_theory}

   In this section, we show how the screened interaction $w^r$ and the polarization $P^r$ can be viewed as the bosonic propagator and self-energy, respectively, of a bosonic theory of an order parameter field $\varphi^r$. In particular, fermion diagrammatics of $w^r$ in the language of the SBE formalism can be consistently mapped to boson diagrammatics of a bosonic theory $S^r[\varphi^r]$. 
   
The first step is to identify the bosonic propagator with the screened interaction  
\begin{align}
\expval{\varphi^{r}_{\sigin\sigin'}(q) \varphi^{r}_{\sigout\sigout'}(q)}_{\text{conn}} := w^{r}_{\sigin\sigin'\sigout\sigout'}(q),\label{eq:bosonic_two_point_func_w}
\end{align}
and the bosonic self-energy with the polarization $P^{r}(q)$ as suggested by Eq.~\eqref{eq:screened_interaction}. The bare bosonic propagator is taken to be the interaction $w_0^{r} = U^{r}$.  

A consistent mapping requires that $P^{r}$, representing the bosonic self-energy, can be expressed as the sum of diagrams built from well-defined bosonic constituents of $S^r[\varphi^r]$. To that end, we study the structure of $P^{r}(q)$. Inserting Eq.~\eqref{eq:U_irreducible_vertex} in Eq.~\eqref{eq:polarization}, we find  
\begin{subequations}
\begin{align}
P^{r}(q) &= \Pi^{r}_{\text{$s$-wave},\text{$s$-wave}}(q) \label{eq:polarization_layedout_bare} \\+ &\int_{k''k'''} \Pi^{r}_{\text{$s$-wave},k''}(q) \mathcal{I}^{r}_{k''k'''}(q) \Pi^{r}_{k''',\text{$s$-wave}}(q), \label{eq:polarization_layedout_correlated}  
\end{align}\label{eq:polarization_layedout}
\end{subequations}
where "$s$-wave" represent integration over the momentum-frequency argument $A_{\text{$s$-wave}} := \int_k A_k$, emphasizing the projection of the momentum dependence to the corresponding $s$-wave component. Disregarding self-energy corrections for the moment, the first term Eq.~\eqref{eq:polarization_layedout_bare}, the ``bare'' polarization, plays the role of a zeroth-order bosonic self-energy, while the second term Eq.~\eqref{eq:polarization_layedout_correlated} gives a ``correlated'' part. Diagrammatically, $\mathcal{I}^r$ sums all (fermionic) Feynman diagrams of the vertex excluding those $U$-reducible in channel $r$, see Eq.~\eqref{eq:U_irreducible_vertex}.

\begin{widetext}

\begin{figure}[h!]
    \centering
\begin{overpic}[width=1.0\linewidth,tics=10]{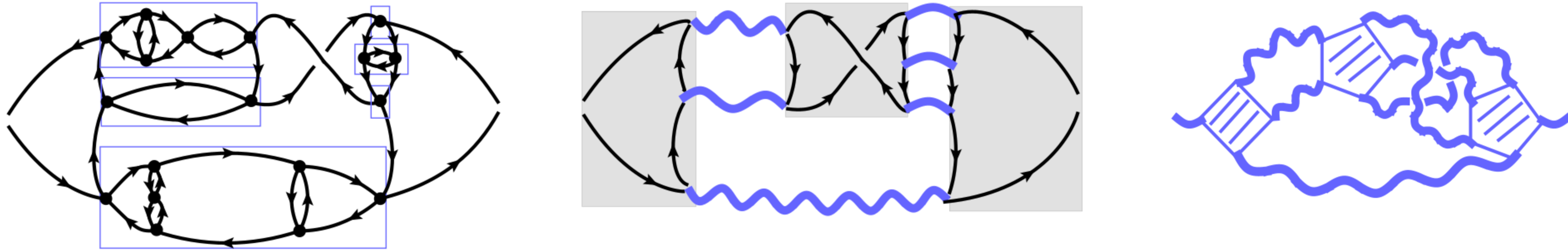}
  \put(-0.7,13){ 1)}
  \put(34,13){ 2)}
  \put(72,13){ 3)}
\end{overpic}
    \\
    \caption{Illustration of diagrammatic bosonization of a polarization diagram. 1) An example of a diagram of the polarization $P^\ph(q)$. The blue rectangles mark contributions to the boson propagators $w^\ph$. 2) Replacing the bosonic propagator contributions with bold full propagators $w^{\ph}$, one obtains a collection of bosonic lines mapped together by electron propagators. In grey, we shade the remaining sub-diagrams; those irreducible with respect to splitting an arbitrary number of bosonic propagators. 3) Identifying the G-polygons with bare bosonic vertices, we arrive at a purely bosonic skeleton self-energy diagram.}
    \label{fig:example_main}
\end{figure}

\end{widetext}

Working diagrammatically, it is possible to identify bosonic constituents in the polarization through the following steps:
\begin{enumerate}
    \item [1.] Select an electronic skeleton diagram; i.e. omitting all electron self-energy correction in its internal lines, of the correlated part of $P^{r}(q)$. Its general structure is 
    \begin{align}
        \int_{k''k'''} \Pi^{r}_{\text{$s$-wave}, k'}A^{r}_{k''k'''} \Pi^{r}_{k''', \text{$s$-wave}},
    \end{align}
    where $A^r$ is a $U$-irreducible diagram in channel $r$.
    \item [2.] Identify in $P^{r}$ all subdiagrams that build up $w^{r}$ (subdiagrams subtended by two $U$ in channel $r$, or a lone $U$) and replace them with wiggly lines (representing bold bosonic propagators).
    \item [3.] The remaining fermionic constituents will be collections of electron propagators that necessarily form $n$-sided $G$-polygons, which we identify with the bare bosonic vertices $\{\mathcal{V}^{(n)}\}$.  
\end{enumerate}
This procedure of "bosonizing" the polarization diagram is unambiguous~\footnote{In applying this procedure to an $r = \ph$ diagram, there is a choice in identifying a bare interaction as a particle-hole boson propagator contribution in a direct or a transverse orientation. The choices available for such a diagram do not represent an ambiguity in the description however, but rather reflect the multiplicity of the fermionic diagram. }, and the result is a pure bosonic skeleton self-energy diagram in terms of bosonic constituents \emph{of a single channel} $r \in \{\ph,\pp\}$. An illustration of these three steps is given in Fig.~\ref{fig:example_main}. 1) Focusing on the $r = \ph$ channel, we select a contribution to the particle-hole polarization $P^\ph$. 2) We identify the bosonic propagator contributions in the $\ph$ channel, marked with blue dotted squares. This diagram is contained in the expansion of a skeleton diagram where these contributions are replaced with full bosonic propagators $w^\ph(q)$. 3) Finally, the remaining $w^{r}$-irreducible components, shaded in grey in the figure, each contain a collection of electron Green's functions that form a 4-sided $G$-polygon and two 5-sided ones. At each vertex of such an $n-$gon, there connects a bosonic propagator. Fixing the convention of the orientation of the loops, they can be identified with bosonic bare vertices of an effective bosonic theory. The result is a purely bosonic skeleton self-energy diagram.

If we do not restrict ourselves to skeleton polarization diagrams, then inserting an electron self-energy correction on an $n$-sided $G$-polygon would effectively turn it into an $(n + m)$-sided polygon. The $m$ additional sides correspond to the external legs of the self-energy insertion, which are then contracted with one another independently of the rest of the diagram. We show this and further worked out examples in  App.~\ref{app:example}. 

Collecting such vertices, two actions can be written down for an $r = \ph$ and an $r = \pp$ theories with G-polygons for interactions 
\begin{widetext}
\begin{subequations}\label{eq:exact_bosonic_theory}
\begin{align}
S^\ph[\varphi^{\ph}] &= \frac{1}{2}\int_{q}\sum_{\sigma_1\sigma_1'\sigma_2\sigma_2'}
\varphi^{\ph}_{\sigma_1\sigma_1'}(q)
\big([U^\ph]^{-1}_{\sigma_1\sigma_1'\sigma_2\sigma_2'}+
\mathcal{V}_{0,\sigma_1\sigma_1'\sigma_2\sigma_2'}^{(2)\ph}(q, -q)\big)
\varphi^{\ph}_{\sigma_1\sigma_1'}(-q) \nonumber \\
&\qquad+ \sum_{n = 3}^\infty \frac{1}{n!}\int_{q_1\cdots q_n}\sum_{\sigma_1\sigma_1'\cdots\sigma_n\sigma_n'}
\mathcal{V}^{(n)\ph}_{0,\sigma_1\sigma_1'\cdots\sigma_n\sigma_n'}(q_1,\ldots,q_n)
\prod_{j=1}^n \varphi^{\ph}_{\sigma_j\sigma_j'}(q_j),
\label{eq:exact_bosonic_theory_ph}
\\[6pt]
S^\pp[\varphi^{\pp}] &= \int_{q}\sum_{\sigma_1\sigma_1'\sigma_2\sigma_2'}
\varphi^{\pp}_{\sigma_1\sigma_1'}(q)
\big([U^\pp]^{-1}_{\sigma_1\sigma_1'\sigma_2\sigma_2'}+
\mathcal{V}_{0,\sigma_1\sigma_1'\sigma_2\sigma_2'}^{(2)\pp}(q, -q)\big)
\varphi^{\pp*}_{\sigma_1\sigma_1'}(q) \nonumber \\
&\qquad+ \sum_{n = 2}^\infty \frac{1}{(n!)^2}\int_{q_1\cdots q_{2n}}\sum_{\sigma_1\sigma_1'\cdots\sigma_{2n}\sigma_{2n'}}
\mathcal{V}^{(2n)\pp}_{0,\sigma_1\sigma_1'\cdots\sigma_{2n}\sigma_{2n}'}(q_1,\ldots,q_{2n})
\Big(\prod_{j=1}^n \varphi^{\pp}_{\sigma_j\sigma_j'}(q_j)\Big)
\Big(\prod_{j=n+1}^{2n} \varphi^{\pp *}_{\sigma_j\sigma_j'}(q_j)\Big),
\label{eq:exact_bosonic_theory_pp}
\end{align}
\end{subequations}
\begin{figure}[h!]
    \centering
    \includegraphics[width=0.9\linewidth]{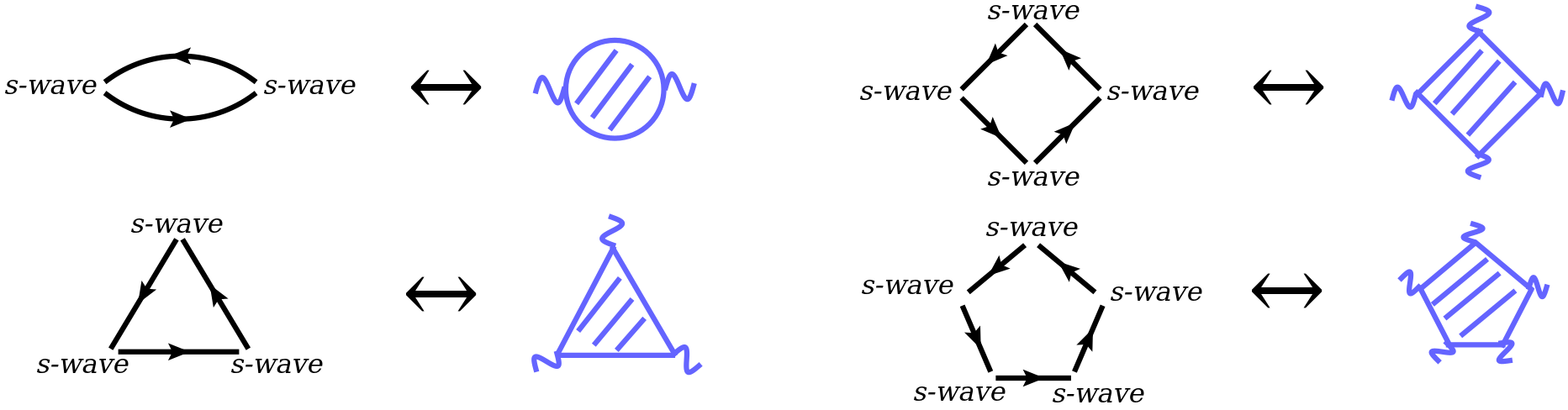}\\
    \vspace{1cm}
    \caption{The 2-point, 3-point, 4-point and 5-point bare bosonic vertices in the $r=\ph$ bosonic theory. The Green's function arrows form a polygon with arrows meeting head-to-tail. Higher-order vertices also appear and are constructed similarly.  The 2-point bare vertex, the Lindhard bubble, plays the role of a zero-order self-energy, that does not appear in the bold/skeleton diagrams. The "$s$-wave" is there to indicate that these are simple loop integrals, and so as bosonic vertices, they only connect to $s$-wave bosonic propagators.}
    \label{fig:natural_bosonic_vertices_sbe_ph}
\end{figure}

\begin{figure}
    \centering
    \includegraphics[width=0.9\linewidth]{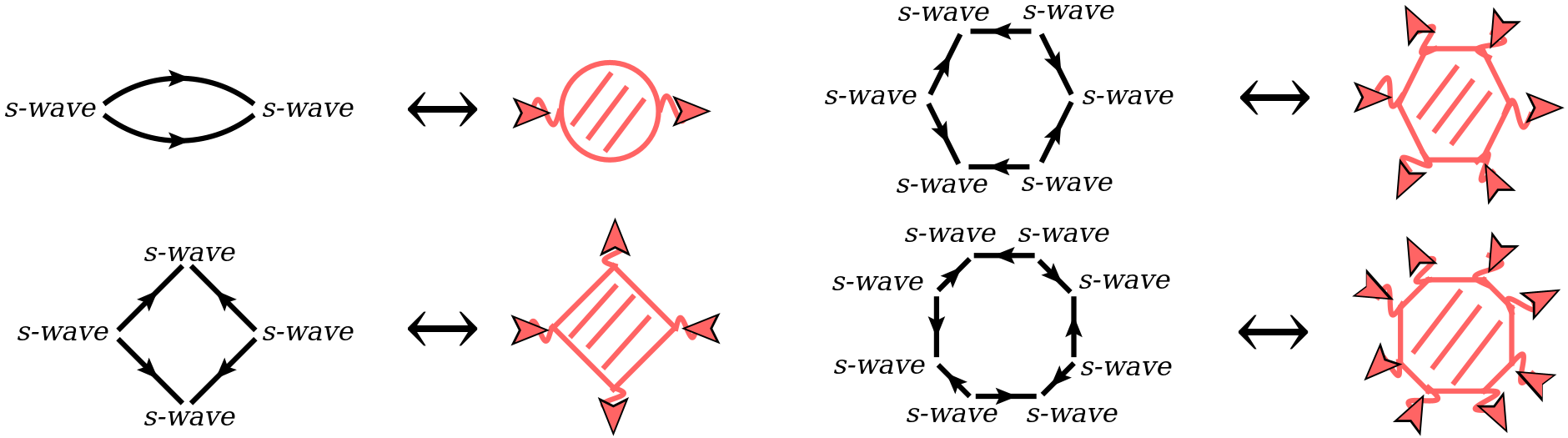}
    \caption{For the $r = \pp$ theory, the vertices are formed of Green's functions whose arrows would instead meet head-to-head and tail-to-tail. Since the theory involves only $\pp$ propagators $w^\pp$, $U(1)$ symmetry entails that only vertices with an equal number of ingoing and outgoing legs can appear.}
    \label{fig:natural_bosonic_vertices_sbe_pp}
\end{figure}

\end{widetext}
for bosonic fields $\varphi^\ph$ and $\varphi^\pp$ respectively. Their primary feature is that their bosonic self-energies and propagators coincide with $P^\ph$ respectively $P^\pp$ and $w^\ph$ respectively $w^\pp$. Notice that only $\pp$ vertices with an even number of legs appear, consistent with $U(1)$-charge symmetry. The $1/n!$ combinatorial factor in the bare vertices of $r = \ph$ theory accounts for the number of ways $n$ bosonic $\ph$ lines can be connected to an $n$-sided polygon. The $1/(n!)^2$ in the $r = \pp$ theory accounts for the number of ways $n$ bosonic \emph{ingoing} lines and $n$ bosonic \emph{outgoing} lines can be connected to their respective corners in a $2n$-sided polygon. The explicit expressions for the $G$-polygon vertices are
\begin{subequations}\label{eq:exact_bosonic_theory_vertices}
\begin{align}
\mathcal{V}^{(n)\ph}_{0,\sigma_1\sigma_1'\cdots \sigma_n \sigma'_n}&(q_1,\cdots q_n) = - \delta(\sum_{j=1}^n q_j)(n-1)! \nonumber \\
\times\int_k &\prod_{j=1}^n G_{0,\sigma_j\sigma_{j+1}'}\left(k + \sum_{l=1}^{j-1} q_l\right), 
\label{eq:Vph} \\
\mathcal{V}^{(2n)\pp}_{0,\sigma_1\sigma_1'\cdots \sigma_{2n} \sigma'_{2n}}&(q_1,\cdots q_{2n}) =\delta(\sum_{j=1}^{2n} (-1)^{j}q_j)n!(n-1)! \nonumber \\
\times \int_k \prod_{\underset{j \leq n}{j\in 2\mathbb{N}}} &G_{0,\sigma'_{j+1}\sigma'_{j}}\left(k - \sum_{l = 1}^{j-1} (-1)^{l}q_l\right) \nonumber\\
 \times 
 \prod_{\underset{j \leq n}{j\in 2\mathbb{N}+1}}&G_{0,\sigma_j\sigma_{j+1}}\left(-k + \sum_{l = 1}^{j-1} (-1)^{l}q_l\right).
\label{eq:Vpp}
\end{align}
\end{subequations}

The combinatorial factor $(n-1)!$ in the $\ph$ vertices is to account for the number of different ways $n$ Green's function lines can form an n-gon. For the $\pp$ vertices, the factor is $n!(n-1)!$, which is the number of ways to connect one column of $n$ elements (charge $+$) to another of $n$ elements (charge $-$). See Figs.~\ref{fig:natural_bosonic_vertices_sbe_ph},~\ref{fig:natural_bosonic_vertices_sbe_pp} for the diagrammatic representation. We stress that there are no vertex functions between the Green's functions in Eqs.~\eqref{eq:exact_bosonic_theory_vertices}, and there are no approximations. As such, the bosonic propagator and bosonic self-energy of these two theories coincide with the SBE objects $w^{\ph}, P^{\ph}$ and $w^{\pp}, P^{\pp}$ respectively. A brief comment on how the discussion above would be modified in the presence of an extended or long-range bare interaction is given in Appendix.~\ref{app:extended_interactions}.

The theories recovered in Eqs.~\eqref{eq:exact_bosonic_theory} bear resemblance to the so-called trace log theory~\cite{HerzKlenin_TraceLogTheory_PhysRevB.10.1084,Herz_QPT_PhysRevB.14.1165,Millis1993QPT_finiteT_PhysRevB.48.7183,Kopietz_1995,Kopietz_1995,Igoshev2007katanin_ferromagnetism_hst}: Traditionally, bosonization of a fermionic theory is achieved by first performing a HST and integrating out the fermions. This is possible if the interaction decouples exactly; i.e., can be written as the square of a bosonic bilinear (up to possible shifts of the chemical potential). Doing so yields a trace log theory. If the interaction doesn't decouple in a single channel, then typically one channel is chosen and the remaining are neglected. Here is how these theories are related to Eqs.~\eqref{eq:exact_bosonic_theory}. Consider the decomposition of the bosonic field into spin-singlet (traceful) and spin-triplet (traceless) parts.
\begin{align}
\varphi^r_{\sigma\sigma'} =  \varphi^{r,0}\delta_{\sigma\sigma'} + \vec{\varphi}^{\, r,1} \cdot \vec{\tau}_{\sigma\sigma'},
\end{align}
with $\vec{\tau}$ the vector of Pauli matrices. For $r = \ph$, this corresponds to the decomposition into charge density ($\D$) and magnetic ($\M$) bosons, while for $r = \pp$ to singlet pairing ($\sSC$) and triplet pairing ($\tSC$) bosons; i.e.
\begin{align}
\varphi^\D &:= \varphi^{\ph,0},\quad \vec{\varphi}^{\,\M} := \vec{\varphi}^{\,\ph,1},\\\varphi^\sSC &:= \varphi^{\pp,0},\quad \vec{\varphi}^{\,\tSC} := \vec{\varphi}^{\,\pp,1} .
\end{align}
Assuming $SU(2)$ invariance of the theory, we have $G_{\sigma\sigma'} = G\delta_{\sigma\sigma'}$. It follows from $(\tau^k)^2 = 1$ and $\Tr \tau^k = 0$ that all terms in the action where the vertices $\mathcal{V}^{(n)}$ are contracted with an odd number of triplet bosonic fields (that is, $\vec{\varphi}^\M$ in the $r = \ph$ theory or $\vec{\varphi}^\tSC$ in the $r = \pp$ theory) vanish - a variant of Furry's theorem. If one neglects charge density fluctuations as is usually done (commonly justified e.g. by being close to half-filling), then our action reduces to the trace log action
\begin{align}
S^\ph[\vec{\varphi}^\M] &\sim \sum_j\frac{1}{n}\Tr[-( G_0\vec{\varphi}^\M\cdot\vec{\tau})^j] \nonumber \\ &= \Tr\ln[1-( G_0\vec{\varphi}^\M\cdot\vec{\tau})],
\end{align}
demonstrating diagrammatically the connection between the bosons in the SBE formalism and the bosons of Hubbard Stratanovich decouplings. A similar conclusion can be reached if one focuses on the $r = \pp$ theory.

Finally, having established a concrete mapping between the diagrammatics of the pure fermionic theory and of the pure bosonic theory Eq.~\eqref{eq:exact_bosonic_theory} diagrammatically enables direct means to study fermionic diagrammatic approximations that utilize the SBE decomposition in the bosonic language. This can be done by: 1) writing down all bosonic diagrams for the bosonic self-energy of the theory (which, by construction, coincide with the polarization $P^r$ in the fermionic language), 2) mapping the bosonic diagrams into fermionic diagrams. 3) Excluding these bosonic diagrams which have that their corresponding fermionic diagrams are not included within the fermionic approximation. 

As an example of where this strategy can be useful, we next use it to revisit a conjectured correspondence between critical PA solutions and the bosonic self-consistent screening approximation (SCSA).

\subsection{Critical parquet approximate solutions and the self-consistent screening approximation}
\label{sec:connection_with_scsa}
In a seminal work by Bickers and Scalapino, it was argued that critical PA solutions where the order parameter has $N$ components lie in the same universality class as the SCSA for the $O(N)$ model. Heuristic arguments were invoked to speculate at the form of a critical order parameter bosonic theory corresponding to a critical PA solution. A selectively integrated generalized susceptibility and a selectively integrated Bethe-Salpeter kernel were identified with the bosonic propagator and self-energies, respectively. Bosonic vertices were not given explicitly, but were assumed at lowest order to be $G$-polygons. The abstract nature of these constructions and lack of an explicit diagrammatic mapping of their bosonic objects to fermionic ones, in our opinion, constitute a gap in their argumentation that has long needed to be addressed. A main conclusion that followed from that work -- that the parquet satisfies the HMW theorem—is therefore also challenged in the same way. 

Armed with the concrete mapping for $w^r$ as a bosonic propagator that we devised in the previous section, we are in a position to revisit this problem, closing these gaps. 

\subsubsection{Power-counting as justification for keeping only the 4th order bosonic interaction}
\label{sec:powercounting}
The presence of an infinite number of bare interactions in the bosonic theory Eq.~\eqref{eq:exact_bosonic_theory} complicates its analysis. To make progress, we note that critical properties are governed solely by the low-energy or long-distance behavior of the theory. 

We thus focus our attention on the low-energy sector of the theory, where, by appealing to power-counting, we will be able to justify neglecting most of the $\mathcal{V}_0^{(n)}$ terms~\cite{Herz_QPT_PhysRevB.14.1165,Millis1993QPT_finiteT_PhysRevB.48.7183}. We may assume without loss of generality a positive anomalous dimension $\eta > 0$, since we will argue later that finite temperature criticality with $\eta \equiv 0$ is excluded.
Furthermore, we consider additionally three extra assumptions: 1) We are approaching criticality in a fixed channel $r^* \in {\pp, \ph}$ with ordering vector $\bfq_*$. 2) A finite temperature transition $T_c > 0$, so that we may neglect effects of Landau damping and dynamical fluctuations. 3) For $r = \ph$ we assume the transition is to a two-sublattice order; i.e., the ordering vector $\bfq_*$ is commensurate with the lattice such that the periodicity of the order parameter on the real lattice is two. This will ensure we can neglect cubic bosonic interactions in the following discussion, which can at most change the nature of the phase transition from second order to first order, a possibility we don't address.

A low-energy theory can be derived by expanding around $\bfq_*$, focusing on the low momentum modes. To that end, the domain of the momentum integral over the Brillouin Zone is replaced by a ball around $\bfq_*$ of radius $\Lambda$ representing a momentum cutoff. The lowest Matsubara frequency is separated from the rest by $\pi T$, so that at low energies and finite temperature only the lowest Matsubara frequency contribution remains in the effective theory, amounting to a theory defined in just $d$-spatial dimensions~\footnote{The standard argument, originally due to Kaluza~\cite{kaluza1921unit} and Klein~\cite{klein1926quantentheorie}, is essentially that compact dimensions (in our case, we deal with thermal time of size $\beta = 1/T$) do not change the low-energy physics in a universal way if its size is small. To make this more precise, let $\varphi(\mathbf{k}, 2\pi T n)$ be a bosonic field, where $n$ here is a bosonic Matsubara index. Viewing the integer $n$ as a label for different species of bosons, i.e., writing $\varphi_n(\mathbf{k})$, then one finds that the bosons $\varphi_n$ with $|n| > 0$ have mass gaps of $(2\pi n T)^2$; and therefore their real space propagators decay exponentially on a scale $\sim 1/(2\pi nT)$. The effect of integrating out modes with $|n| > 1$ on the effective m-point couplings can be found by summing up all Feynman diagrams with $m$ external legs and intermediate $\varphi_n$ propagators with $|n| > 1$. At finite temperature $T$, these contributions should be exponentially suppressed.} Furthermore, we assume our bosonic theory is invariant under $(\bfq-\bfq_*) \rightarrow -(\bfq-\bfq_*)$. The quadratic part of the theory is given by 
\begin{widetext}
    \begin{align}
S_0^{\Lambda}[\tilde{\varphi}] = \int_{\bfq-\bfq_*}^\Lambda \tilde{\varphi}(-\bfq) \left(U^{-1} + \Pi_{\Lambda}(\bfq_*) + (\bfq - \bfq_*)^2\partial^2\Pi_{\Lambda}(\bfq_*) + \mathcal{O}((\bfq - \bfq_*)^4)\right) \tilde{\varphi}(\bfq),
\end{align}
\end{widetext}
where the transfer momentum integral is cutoff by a momentum scale set by $\Lambda$, and contains implicitly in the notation an additional factor of $T$ (the normalization factor of the neglected Matsubara frequency summation).
Power-counting with respect to the free kinetic term $\sim \tilde{\varphi} (\bfq-\bfq_*)^2 \tilde{\varphi}$ gives that under a scaling $\Lambda \rightarrow s\Lambda$ with $0 \ll s < 1$, the field is scaled by $\tilde{\varphi} \rightarrow s^{-\frac{d+2 - \eta}{2} } \tilde{\varphi}$, where we have introduced the anomalous dimension $\eta$ characteristic of the critical point. An operator with $n_{\varphi}$ multiples of $\tilde\varphi$ and $n_{\tilde\varphi}-1$ integrals with $n_\bfq$ (even) factors of momentum $\bfq$ will thus be scaled by $s^{(n_{\tilde\varphi} - 1)d + n_{\bfq} - n_{\tilde\varphi}\frac{2+d - \eta}{2}  }$
. In particular, for the quadratic terms $n_{\tilde\varphi} = 2$, the terms scale as $s^{n_\bfq - 2 +\eta}$, irrespective of the dimension $d$. Therefore, power-counting gives us that the $\mathcal{O}(\bfq^4)$ terms in the expansion of the quadratic part of the action are irrelevant. For higher-point interactions, the dimension plays a role. In $d = 3$, operators scale by $s^{n_{\tilde\varphi}(1+\eta)/2 - 3 + n_\bfq}$, therefore for all $n_{\tilde\varphi} \geq 3$, terms with $n_{\bfq} \geq 2$ are irrelevant. All in all, for $d \geq 3$, the effective theory is given by
\begin{widetext}
\begin{align}
S_{\Lambda}[\varphi] =& \int_{\bfq-\bfq_*}^\Lambda \tilde{\varphi}(-\bfq) \left(U^{-1} + \Pi_{\Lambda}(\bfq_*) + (\bfq - \bfq_*)^2\partial^2\Pi_{\Lambda}(\bfq_*)\right) \tilde{\varphi}(\bfq)\\& + \mathcal{V}^{(4)}_{\Lambda}(\bfq_*, \bfq_*, \bfq_*, \bfq_*)\int_{\bfq_1, \bfq_2, \bfq_3}^\Lambda \tilde{\varphi}(\bfq_1)\tilde{\varphi}(\bfq_2)\tilde{\varphi}(\bfq_3)\tilde{\varphi}(-\bfq_1 + \bfq_2 - \bfq_3), \label{eq:minimal_effective_action_near_criticality}
\end{align}
\end{widetext}
where the $3-$point interaction is absent for $r_* = \pp$ by $U(1)-$charge symmetry prohibiting any $n_{\tilde\varphi}$ odd, and for $r_* = \ph,\xph$ by assumption 3), which makes it impossible for an odd number of momenta $\bfq_i$ to be $\Lambda$-close to $\bfq_*$ while simultaneously conserving momentum. Splitting the field into a singlet and a triplet component yields a quartic $O(N)$ theory with $N = 1$ for the singlet order parameter, and $N = 3$ for the triplet one for the particle-hole $r_* = \ph$ case. For the particle-particle $r_* = \pp$ case, we have $N = 2$ and $N = 6$ corresponding to the number of components of singlet-pairing and triplet-pairing order parameters, respectively.

The situation for $d \equiv 2$ is more subtle, as naive power-counting as done above leads to an arbitrary number of relevant operators for arbitrarily small $\eta$. It turns out that $\tilde{\varphi}$ in this situation is not an eigen-operator under scaling~\cite{mussardo2010statistical,cardy1996scaling}, and therefore power-counting is not justified. Here, conformal field theory arguments can be invoked~\cite{minimalmodels_BPZ1984,mussardo2010statistical} for the specific case $N = 1$ to nonetheless justify neglecting sextic and higher-order interactions, and we end up again with a theory given by Eq.~\eqref{eq:minimal_effective_action_near_criticality}~\footnote{This is the same reason Ising universality can be represented by the Ising model that does not contain sextic and higher-order interactions.}.

The above arguments tell us that to obtain the correct universal physics for the action Eq.~\eqref{eq:exact_bosonic_theory} near a second-order transition at $d \geq 3$ with any $N$ or $d = 2$ with $N = 1$ with the assumptions stated at the beginning of this section, the only relevant terms are: 1- the static quartic interaction $V_0^{(4)}$ evaluated at ordering vectors $\bfq_*$ leaving us with a quartic coupling constant, and 2- the static quadratic part expanded to second order in deviations from $\bfq_*$. This yields a mass term (from the zeroth-order bubble and the inverse bare interaction) and a stiffness term (from the second derivative of the bubble).

\subsubsection{The critical parquet approximation in the bosonic language}
By enumerating through the relevant bosonic diagrams, we may determine which of them correspond to the PA in the bosonic language.

The strategy is to investigate the contributions of the bosonic self-energy (corresponding to the polarization $P^{r_*}$) of the theory Eq.\eqref{eq:minimal_effective_action_near_criticality} diagrammatically and then translate these contributions to the fermionic language. The diagram is only included if it is a diagram generated within the fermionic PA. Recall that a PA vertex diagram (henceforth "PA diagram" for short) can be identified inductively from the following rules: 1) The bare interaction diagram is a PA diagram. 2) $GG$-reducible diagrams: If $A$ and $B$ constitute PA diagrams, then so does $\int_{kk'}A^r_{kk'}\Pi^r_{kk'} B^r_{kk'}$ with $r = \pp, \ph, \xph$ (you can cut a $GG$ pair to split it into $A$ and $B$). From this definition, a fermionic vertex Feynman diagram that is completely $GG$-irreducible in all three channels is not a PA diagram (unless it is the bare interaction). Moreover, a diagram that happens to be $GG$-reducible, but either or both of $A$ and $B$ that result from cutting the $GG$ pair are $GG$-irreducible, is not a PA diagram. A diagram is only a PA diagram if $A$ and $B$ are both $GG$-reducible (or a bare interaction), and each further split into $GG$-reducible (or bare interaction) diagrams. This is required \emph{ad infinitum}.

\begin{figure}[h!]
        \centering
        \includegraphics[width=0.8\linewidth]{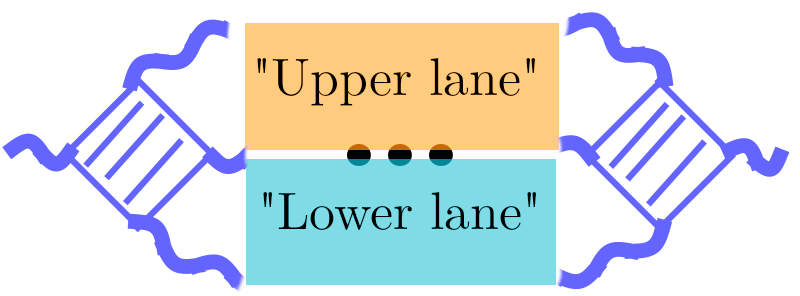}
        \caption{In the diagrammatic analysis for $\mathcal{V}_0^3$ and higher skeleton self-energy diagrams in the main text, we refer to by ``upper lane'' and ``lower lane'' the spatial regions marked in this figure. }
        \label{fig:upper_lower_lane}
    \end{figure}

By construction, the bosonic propagator $w^{r_*}$ constitutes a parquet solution if it is built from a PA polarization $P^{r_*}$. Thus, it is sufficient to analyze $P^{r_*}$. We proceed order-by-order in $\mathcal{V}_0^{(4)}$ -- the only relevant bare interaction -- which we denote by $\mathcal{V}_0$ for brevity. We need to only focus on the skeleton diagrams of $P^{r_*}$ -- where we assume $w^{r_*}$ appearing there are fully dressed by the same $P^{r_*}$. The correlated part of the polarization has the structure $\Pi \circ \mathcal{I} \circ \Pi$ where $\mathcal{I}$ is a vertex; see Eq.~\eqref{eq:polarization_layedout}. Thus, in doing the analysis, we must ensure to exclude the two external $GG$ pairs. To simplify the analysis, we will only consider a single permutation of the bosonic vertex corresponding to ``square''; a seemingly innocuous simplification but is subtle; see Appendix.~\ref{app:permutations_considerations}. For third- and higher-order skeleton bosonic self-energy diagrams, we will refer to the "upper lane" and "lower lane" parts of the diagram to reference spatially the regions shown in Fig.~\ref{fig:upper_lower_lane}. 

\begin{figure}[h!]
    \centering
    \includegraphics[width=0.7\linewidth]{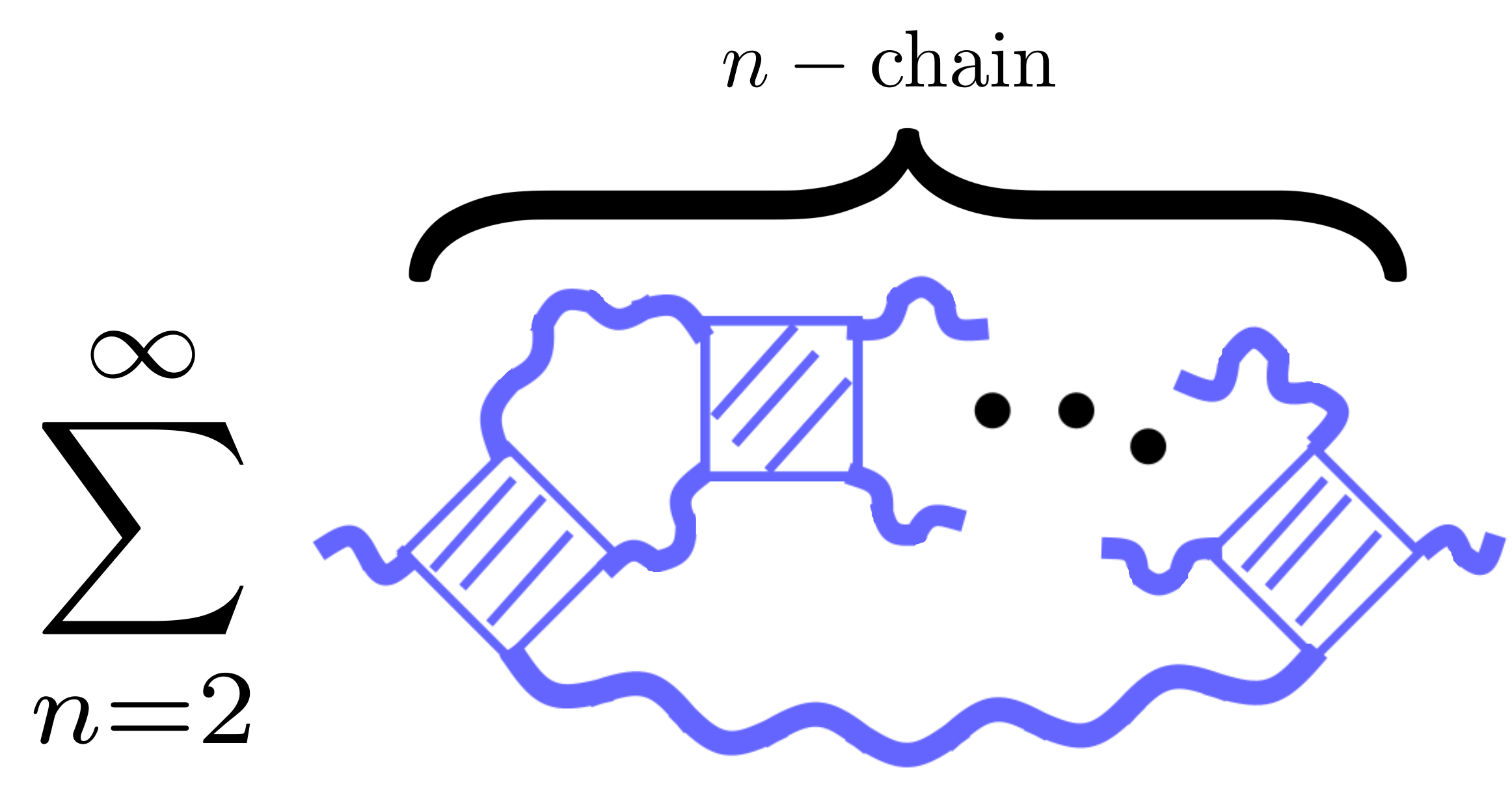}
    \caption{Summation of all relevant correlated skeleton bosonic self-energy diagrams in the order parameter theory of the PA near criticality.}
    \label{fig:rainbow_summation}
\end{figure}

\begin{figure}[h!]
    \centering
    \includegraphics[width=0.85\linewidth]{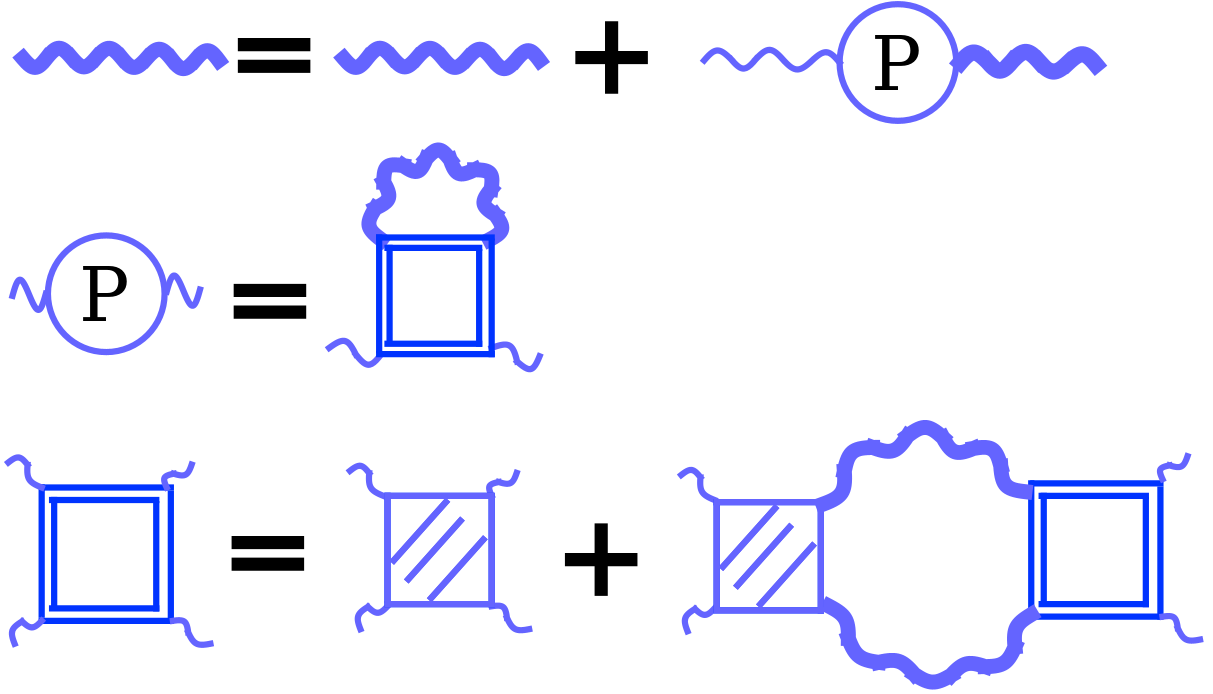}
    \caption{Self-Consistent Screening Approximation for the $O(N)$ models as a self-consistent version of the large $N$ approximation. The plain (bold) wiggly lines represent the bare (dressed) bosonic propagators, and the hashed square (bold square) represents the bare (dressed) bosonic 4-point interaction. The equations are solved -- upto the tadpole term -- by taking for the bosonic self-energy, which we here denote by P, the rainbow diagrams as shown in Fig.~\ref{fig:rainbow_summation}. The bosonic propagators are self-consistently dressed with P using the Dyson equation (first line). }
    \label{fig:scsa_original}
\end{figure}

\begin{enumerate}[label={}, leftmargin=0pt]
    \item \textbf{First-order}:  This is simply the ``tadpole diagram''. By translating this diagram to the fermionic language, it is seen to be a bare bubble contribution to the polarization diagram where one of the fermion Green's function has an explicit self-energy correction; see Fig.~\ref{fig:hartree-fock_boson}. Therefore, this diagram is not a skeleton in the fermionic propagators, and including it would amount to a shift of the electronic self-energy. This can be treated by introducing a counter ``bare'' fermion self-energy.  
    \begin{figure}[h!]
\centering
         \includegraphics[width=0.9\linewidth]{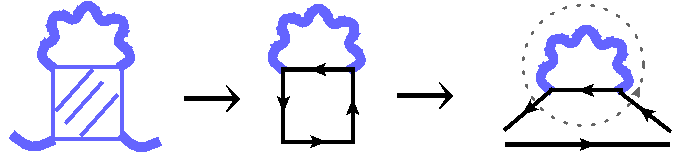}
        \caption{The first-order ``tadpole'' diagram of the bosonic theory is translated by replacing the box with the square of Green's function lines. It corresponds to a bubble where one of the electronic lines is renormalized by a screened interaction and will thus effectively result in a mere shift of the value of the electron self-energy dressing the bold electron propagators. }
        \label{fig:hartree-fock_boson}
    \end{figure}
    \item \textbf{Second- and third- order}:  There is only one topologically distinct diagram for each order $\mathrm{O}(\mathcal{V}^2_0)$ and $\mathrm{O}(\mathcal{V}^3_0)$; the ``sunset diagram'' and the "rainbow diagram" shown in Fig.~\ref{fig:second_and_third_order}. In translating these diagrams to the fermionic language, one finds both to be PA contributions. After removing the two external $GG$ pairs, it can be verified explicitly that the remaining are indeed PA diagrams. This is done analogously to what is shown for the fourth-order rainbow diagram shown in Fig.~\ref{fig:fourth_order_pa_not_pa} a). After repeatedly splitting the diagram into twos, the result will be a collection of bosonic propagators, each of which is, by hypothesis, an interaction dressed by a PA polarization. Therefore, the two diagrams are included within the PA. 
    \begin{figure}[h!]
        \centering
        a) \includegraphics[width=0.33\linewidth]{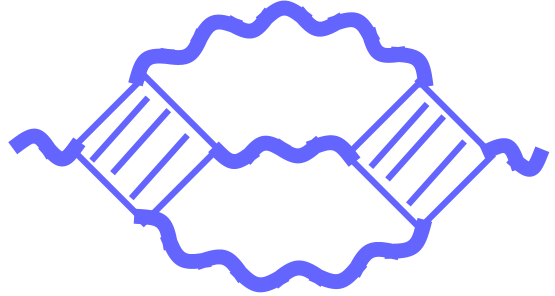}\hspace{1cm}
        b) \includegraphics[width=0.33\linewidth]{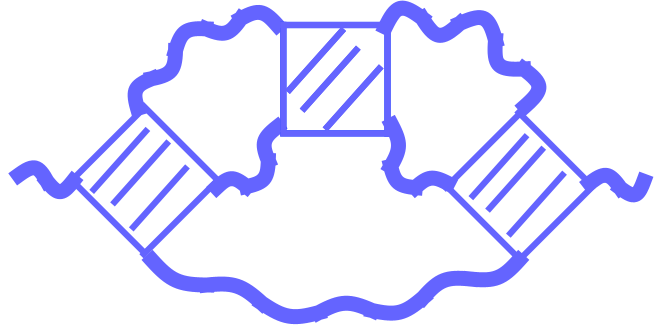}
        \caption{Second and third order skeleton contributions to $P^{r_*}$. After replacing the bare bosonic vertices with $G$-squares, it can be readily verified that both of these constitute parquet diagrams. See also Fig.~\ref{fig:fourth_order_pa_not_pa}a). }
        \label{fig:second_and_third_order}
    \end{figure}
    \item \textbf{Fourth-order}: There are three topologically inequivalent skeleton diagrams at $\mathrm{O}(\mathcal{V}_0^4)$, shown in Fig.~\ref{fig:4thorder}. The first is a rainbow diagram and is readily shown to be a PA diagram. The second diagram, on the other hand, deviates from being a rainbow in that the third boson interaction vertex connects the upper lane to the lower lane. This diagram can be seen to not be in the PA and not even -- after removing the two $GG$ pairs in the definition of the correlated part of the polarization -- $GG$ -- $GG$-reducible. The third diagram deviates from being a rainbow diagram in that it introduces a "vertical" structure in the upper lane. Since this diagram keeps the upper lane and the bosonic propagator on the lower lane separate, it is found to be $GG$-reducible. However, the sub-diagram that results on the upper lane after cutting the $GG$ pair is not $GG$-reducible, which can be attributed to this vertical insertion. Therefore, the third diagram is not a PA diagram.
\begin{figure}[h!]
    \centering
    \begin{overpic}[width=1.0\linewidth,tics=10]{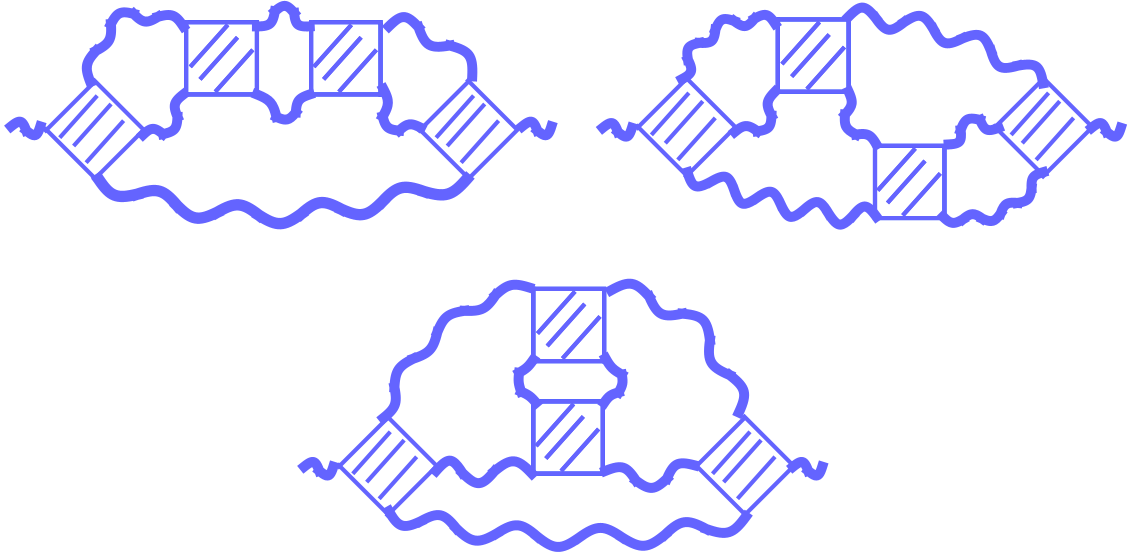}
  \put(-0.1,48){ a)}
  \put(52,48){ b)}
  \put(27,20){ c)}
\end{overpic}
    \caption{The three topologically distinct fourth-order skeleton self-energy diagrams. a) A rainbow diagram. b) A diagram where the two lanes are interconnected. c) A diagram with a vertical insertion in the upper lane.   }
    \label{fig:4thorder}
\end{figure}
    
     \begin{figure}[h!]
        \centering

        \begin{overpic}[width=1.0\linewidth,tics=10]{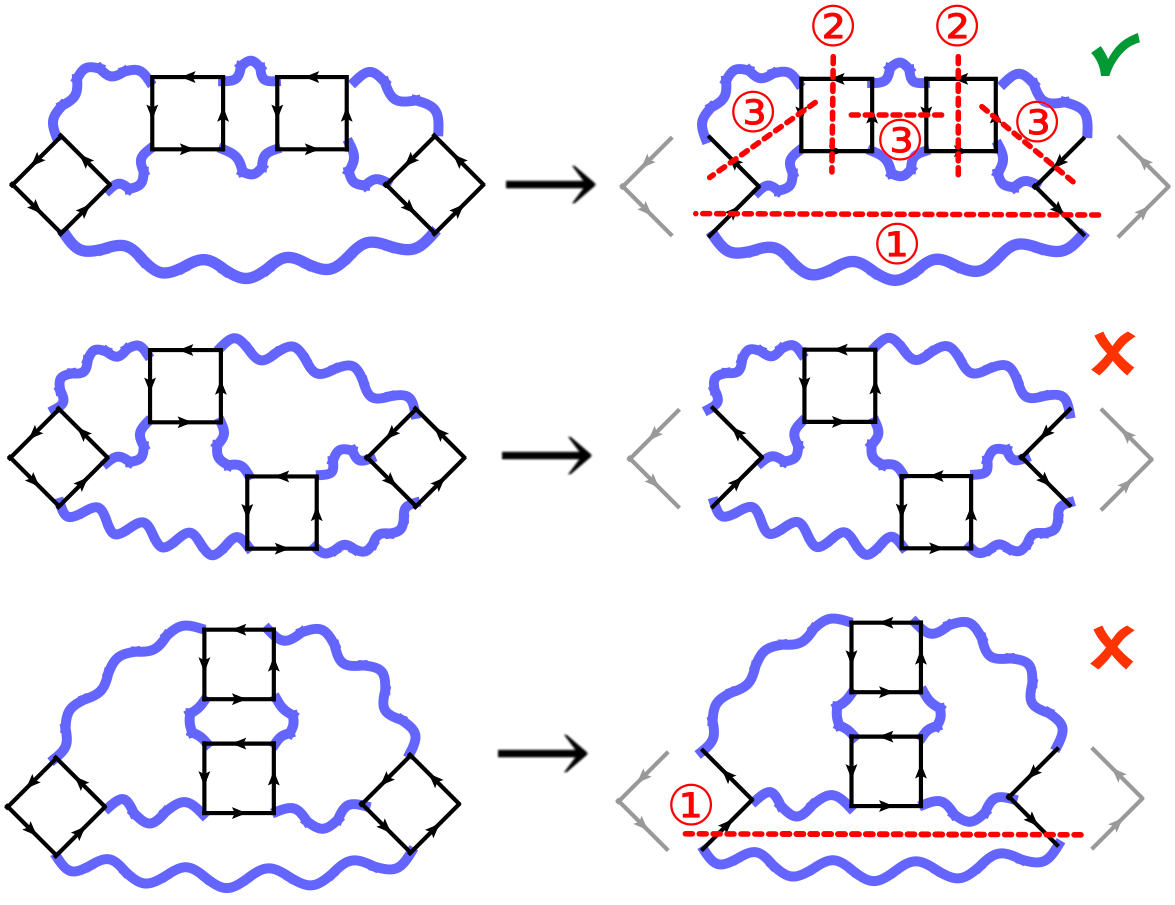}
  \put(-0.1,72){ a)}
  \put(-0.1,48){ b)}
  \put(-0.1,20){ c)}
\end{overpic}
        \caption{The three diagrams of Fig.~\ref{fig:4thorder} are analyzed from the fermionic viewpoint. The two external $GG$ pairs shown in grey are always identified with the bubbles in Eq.~\eqref{eq:polarization_layedout_correlated} and removed, while the remaining is tested for being a PA vertex diagram.   In a), it is shown how "rainbow" diagrams belong to the correlation part of the PA polarization. To see this, note that one can start by cutting the $GG$ pair labeled by \ding{172}, which splits the diagram into a bosonic propagator (a PA vertex diagram) on the lower lane and an upper piece. To see the upper piece is also a PA vertex diagram, one can cut any of the pairs labeled \ding{173} to split the diagram into two parts. The propagator pairs labeled \ding{174} can then be cut, with each cut splitting a subdiagram into two, until eventually one is left with a collection of bosonic propagators. In b), we see that after removing the two bubbles, there is no $GG$ pair that can split the diagram into two; thus, this diagram is not a PA diagram. In c), we see that the diagram is $GG$-reducible, but the resulting subdiagram in the upper lane is not. Thus, c) is also not included in the PA polarization.  }
        \label{fig:fourth_order_pa_not_pa}
    \end{figure}
     \item \textbf{Fifth-order (and higher)}:  There will be a variety of topologically distinct skeleton diagrams, some of which will be rainbow, and can be explicitly verified to constitute PA diagrams, completely analogously to what is shown in Fig.~\ref{fig:fourth_order_pa_not_pa}a). The others deviate from being rainbow in that they connect the upper and lower lanes as in Fig.~\ref{fig:4thorder}b) and/or that they have vertical insertions in the upper or lower lanes as in Fig.~\ref{fig:4thorder}c). This excludes them from being PA diagrams.

\end{enumerate}
Thus, the result of this discussion is that in the bosonic language, the critical PA solution has a skeleton bosonic self-energy (polarization) produced by summing up all the rainbow diagrams, see Fig.~\ref{fig:rainbow_summation}, with the bosonic propagators appearing in the summation self-consistently dressed with this bosonic self-energy. Considerations of other permutations of fermionic lines making up the fermion vertex are discussed in Appendix.~\ref{app:permutations_considerations} and are found to not change this conclusion in a crucial way (only the first-order tadpole and second-order rainbow are affected). This summation happens to be the solutions of the SCSA equations, shown diagrammatically in Fig.~\ref{fig:scsa_original}. This approximation resembles a bosonic version of $GW$ (not to be conflated with the fermionic GW presented at the end of Section~\ref{sec:sbe_intro}). It is also described as a self-consistent large-N. The conclusion is that the critical exponents of the PA, where the order parameter has $N$-components, are the same as the critical exponents of the SCSA for an $O(N)$ theory. In particular, the anomalous dimensions are $\eta_{\text{PA}} = \eta_{\text{SCSA}}(d, N)$, where for $d > 2$ we have that $N$ can be any number. Whereas in $d \equiv 2$, the result strictly holds only for $N = 1$. Although, if one assumes the ability to analytically continue $\eta_{\text{PA}}(d = 2^+, N) = \eta_{\text{PA}}(d = 2, N)$, then one may generally assume that $\eta_{\text{PA}} = \eta_{\text{SCSA}}$ also at $d = 2$ for any $N$. 

In the seminal work~\cite{bray_scsa_original_1974}, Bray employed a bootstrap method to calculate the anomalous dimensions of the $O(N)$ model within the SCSA. The results are as follows. For $d = 3$, it was found that $\eta_{\text{SCSA}}(d = 3, N = 1) = 0.177, \eta_{\text{SCSA}}(d = 3, N = 2) = 0.11$, and $\eta_{\text{SCSA}}(d = 3, N = 3) = 0.079$. For $d = 2$, it was found that $\eta_{\text{SCSA}}(d = 2, N > 1) = 0$ vanishes for any $N > 1$. For $N = 1$, it was found that either $\eta = 0$ or $\eta = 0.453$ are possible. To our knowledge and understanding, it has not been determined which. The case for $N = 1$ corresponds to an Ising-like order parameter; in the electronic problem, this is the order parameter of a doubly degenerate CDW. 

We comment briefly on the SBE(b) approximation, which truncates on the PA by neglecting the rest functions $M^r$. The analysis above for the PA can be straightforwardly repeated by additionally requiring the PA diagrams to be $U$-reducible. It is then found that all rainbow diagrams of order $\mathcal{V}_0^3$ and above are excluded. Thus, the SBE(b)'s critical exponents are expected to simply be those of mean-field; i.e., $\eta_{\text{SBE(b)}} = 0$. Finally, we believe that full D$\Gamma$A or any real space cluster extension of these methods -- which approximates $V_{\text{2PI}}$ by a local quantity, or one on a finite cluster, cannot change these finite temperature universal aspects. Our reasoning for this is that since frequency dependence is irrelevant, introducing a local or a cluster $V_{\text{2PI}}$ amounts to changing the value of the fermionic bare interaction either: by a local value in a local approximation of $V_{\text{2PI}}$, or by a finite-range function for a real-space cluster approximation.

\section{Hohenberg-Mermin-Wagner Theorem}
\label{sec:mermin_wagner_with_self_energy}

\subsection{Role of the self-energy} 
A claim first made in \cite{parquet_critical_bickers_scalapino_1992} is that a finite self-energy forbids finite temperature transitions characterized by a vanishing exponent $\eta$ in two dimensions. In this section, we substantiate this claim and show that: i) if we are in $d = 2$, ii) there is a second order transition with a vanishing anomalous dimension $\eta = 0$ and iii) the approximate state ($\Sigma$, $V$) is converged at the level of the self-energy and the vertex, then the critical temperature associated to that transition must vanish $T_c = 0$, precluding a finite temperature second order transition. We will show a slightly stronger statement, that the finite temperature transition is precluded if $\eta + d \leq 2$.

The key observation needed to justify this claim is to note that there is a negative feedback loop present between the self-energy and bosonic fluctuations $w^r$~\cite{pseudogapFLEX_PhysRevLett.76.1312,pseudogap_tpsc_Vilk_1997,weak_vs_strong_pseudogap_PhysRevLett.80.3839,dampingspectralfunc_PhysRevB.77.115129,pseudogap_Krien_2022,pseudogap_SDE_PhysRevResearch.2.033068}. Suppose we converge the vertex cycle (light grey box in Fig.~\ref{fig:solution_strategy}) with an initial $\Sigma = 0$ to obtain a bosonic propagator $w_0^{r}$. Updating the self-energy with Eq.~\eqref{eq:SDE_SBE} yields a finite self-energy $||\Sigma_{1}|| > 0$. This self-energy generally suppresses the bubbles $\Pi^r$, which in turn dampens the polarization $P^{r}_1$ (see Eq.~\eqref{eq:polarization}). Consequently, the screened interaction $||w^r_{1}||$ is reduced relative to $||w^r_{0}||$ (see Eq.~\eqref{eq:screened_interaction}). In the subsequent self-energy iteration, this smaller $w^r_1$ produces a smaller self-energy $||\Sigma_2|| \leq ||\Sigma_1||$, which would then allow $w^r$ to increase. This iterative feedback continues until $w^r$ and $\Sigma$ reach a set point equilibrium.

To see how this feedback prevents criticality in low dimensions, consider a scenario of a converged solution at the level of the self-energy and the vertex (assumption iii))  at a finite temperature $T > 0$ that is moreover critical. For concreteness, but without loss of generality, we assume a magnetic instability ($\X = \M$). 
This implies a diverging susceptibility near the ordering vector as
\begin{align}
\chi^{\M}(i\Omega = 0, \bfq_* + \delta \bfq) &\sim \frac{1}{\abs{\delta \bfq}^{2-\eta}}\label{eq:scaling_relation}.
\end{align}
This singular behavior is inherited by the screened interaction $w^{\M}$ since in general
\begin{align}
    \chi^\M = (U^\M)^{-1}(U^\M - w^\M) (U^\M)^{-1}. \label{eq:susceptibilities_from_w_diagrammatic}
\end{align}

To make the feedback of the critical magnetic fluctuations on the self-energy explicit, we utilize the $\M$-channel form of the SDE~\cite{Patricolo_2025}
\begin{align}
\Sigma(k) = \int_{q} w^\M(q)\lambda^\M_{k-q}(q)G(k-q). 
\end{align}
Writing $q = q_* + \delta q$ with $q_* = (i0, \bfq_*)$, the integration range can be split into a part close to the ordering vector and at zero frequency $\delta q < \epsilon$ ($i\Omega = 0$ and $\norm{\bfq} < \epsilon$) where the form Eq.~\eqref{eq:scaling_relation} applies, and a part including the rest $\delta q \geq \epsilon$ (and $i\Omega \neq 0$). We have
\begin{align}
    \Sigma(k) &=  \int_{\delta q \geq \epsilon}  w^\M(q) \lambda^\M_{k-q}(q) G(q - k) + \label{eq:split_sig_integrand}\\  &\hspace{18mm}\int_{\delta q < \epsilon}  w^\M(q)\lambda^\M_{k-q}(q) G(q - k)  \nonumber \\
&\approx   \Sigma_{regular}(k) + \int_{\abs{\delta \bfq} < \epsilon} \hspace{-5mm}\frac{G(-i\nu, \bfq_* - \bfk)}{ \abs{\delta \bfq}^{2-\eta}} \label{eq:sig_critical_expr}.
\end{align}
In arriving to the final expression we have assumed the Hedin vertex can be approximated by its bare value $\lambda^\M \approx 1$ for our discussion. See Appendix.~\ref{app:alternative_se_proof} for an alternative argumentation that that does not assume this. The first term in the final expression is regular, whereas the regularity of the second term depends on the dimension $d$ and the value of $\eta$ associated to this instability. If $\eta + d > 2$, the integral converges and the self-energy is regular for general $k$ -- nothing unusual. If on the other hand, $\eta + d \leq 2$, then the second term diverges, leading to a divergent self-energy $\Sigma(k)$ for almost all values of $\bfk$~\footnote{If $G(-i\nu, \bfq_* - \bfk)$ is zero, which can happen in a Mott insulating phase (at least at very low temperatures), then $\Sigma(i\nu, \bfk)$ may be finite even if $\int |\bfq_*|^{\eta-2}$ is divergent. Excluding limits which alter the dimensionality of the problem (e.g. atomic limit, limit of infinite dimensions), the Green's function zeroes would form a surface of codimension 1. Thus, we wrote ``almost all $\bfk$'' to be careful.}. In the marginal case $d = 2,\eta =0$ of assumptions i) and ii), the divergence is logarithmic. A self-energy that diverges almost everywhere on a finite volume region in momentum space is nonphysical, and further does not constitute a converged solution of the parquet equations. This is in contradiction to assumption $iii)$, that the solution we started with is already converged at the level of both the self-energy and the vertex.

Assumptions i) and ii) are relevant in the discussion of the antiferromagnetic instability ($\bfq_* = (\pi, \pi)$) in the Hubbard model on the square lattice ($d = 2$), as the magnetic order parameter has $N = 3$ components, and within the PA $\eta(d=2, N=3) = 0$. We note at this point that PA results that neglected the self-energy and exhibit an antiferromagnetic instability has been reported in \cite{Eckhardt_2018}. Neglecting the self-energy amounts to fixing $\Sigma = 0$ (which is only plausible to be a a part of a fully converged solution at $U = 0$), and thus such a solution is converged only at the level of the vertex. If one were to attempt to update the self-energy with the SDE, one finds that the system is driven out of the equilibrium by a diverging self-energy that suppresses the diverging $\chi^\M$\footnote{
In practical numerical calculations on a computer, operations involving extremely large and extremely small numbers cannot be represented accurately at the same time. As a consequence, it is not possible to work exactly at the critical temperature, and one must content themselves with working at a temperature slightly larger than the theoretical critical temperature. In this context, a quantity such as $w^\M$ or $\Sigma$ being divergent should be understood as being very large (but finite) within machine precision, while vanishing should be understood as being very small (but finite) within machine precision.} in the next vertex cycle. This illustrates how feedback from a large self-energy drives one away from criticality, effectively reducing the critical temperature $T_c$. Subsequent calculations within the same group~\cite{Eckhardt_2020} observed that when the PA solution is converged also at the level of the self-energy, divergence of the antiferromagnetic susceptibility was delayed, consistent with our discussion.

Since a consistent self-energy must diverges for almost all $\bfk$ regardless of the temperature $T > 0$, the critical temperature is pushed towards zero. As we approach the vanishing $T_c$ however, dynamical quantum effects like Landau damping become relevant increasing the effective dimensionality of the problem. This modifies the scaling behavior of $w^r$ at criticality and may alter the above conclusion on the magnetic transition in the Hubbard model. This is consistent with the HMW theorem, which does not preclude a zero temperature magnetic transition.

\subsection{Role of crossing-symmetry}
Another property of a many body approximation which has been argued to lead to the fulfillment of the HMW theorem is the crossing symmetry of the vertex
\begin{align}
V(k_1, k_2, k_3, &k_4)_{\sigma_1\sigma_2\sigma_3\sigma_4} \nonumber \\  =-V(k_1, &k_4, k_3, k_2)_{\sigma_1\sigma_4\sigma_3\sigma_2} \nonumber \\  =-V(k_3, &k_2, k_1, k_4)_{\sigma_3\sigma_2\sigma_1\sigma_4} \nonumber \\  =V(k_3, &k_4, k_1, k_2)_{\sigma_3\sigma_4\sigma_1\sigma_2}, \label{eq:crossing_symmetry}
\end{align}
which follow from the anticommuting nature of the fermionic Grassmann variables. A parquet vertex that solves the parquet equations with $V_{\text{2PI}}$ satisfying Eq.~\eqref{eq:crossing_symmetry} is itself crossing symmetric~\cite{Bickers2004bookchapter}. Since $V_0$ in Eq.~\eqref{eq:bare_vertex_spin_invt} satisfies Eq.~\eqref{eq:crossing_symmetry}, the PA -- with $V_{\text{2PI}} = V_0$ -- is therefore crossing symmetric as well. As a consequence of this fact, the following local sum rule holds~\cite{chalupa_sumrules}
\begin{align}
\frac{1}{2}\sum_q (\chi^\M_{\text{correl}}(q) + \chi^\D_\text{correl}(q))  = 0,\label{eq:local_sumrule}
\end{align}
where $\chi_{\text{correl}}^\X = \chi^\X - \Pi^\X$ the correlated part of the susceptibilities. In two dimensions, for the AFM pseudo-critical transition with a singular behaviour $\chi^\M \sim \abs{\bfq - \bfq_*}^{2-\eta}$ and $\eta = 0$, we find ourselves once again in a situation the momentum integral diverges logarithmically; in this case, this would lead to a violation of the sum rule necessitated by crossing symmetry. Thus, it has been claimed that fulfillment of a sum rule like Eq.~\eqref{eq:local_sumrule} implies the fulfillment of the HMW theorem. An important distinction to the situation of the self-energy of the previous section is that Eq.~\eqref{eq:local_sumrule} is not imposed directly, but is a consequence of another fact, namely the crossing symmetry. In practice, state of the art calculations of the parquet equations in the thermodynamic limit lead to small violations of the crossing symmetry due to  sensitivity in the treatment of momentum dependence~\footnote{See App.~D of \cite{Eckhardt_2020}.}. Thus, by itself, formal satisfaction of Eq.~\eqref{eq:local_sumrule} is a less robust indicator that finite temperature transitions in $d = 2$ are forbidden in numerical simulations, compared to self-consistency with the self-energy which involves a negative feedback loop.  
In contrast, several approximations such as the Two Particle Self-Consistent~\cite{TPSC_original_PhysRevB.49.13267,TPSC_rev_Tremblay_2011,TPSC_delre2025twoparticleselfconsistentapproachbroken,TPSC_jonas_10.21468/SciPostPhys.19.1.026} (TPSC) approach and the $\lambda$-corrected D$\Gamma$A~\cite{lDGammaAKatanin_2009} enforce the local sum rule directly by introducing a negative feedback loop between the growth of the vertex and the left hand side of the sum rule. Such approximations then robustly satisfy the HMW theorem, regardless of a self-consistent self-energy.

\subsection{Numerical fit for $\eta$ in the parquet approximation}
In this section we present a brute force attempt at calculating the anomalous dimension $\eta$ for a CDW ($N = 1$) and AFM ($N = 3$) for the Hubbard model on the square lattice, where, in order to see criticality even for $\eta = 0$, we neglected the self-energy. To that end we have also neglected frequency dependence of the vertices, which should not change the universal physics as discussed in Section.~\ref{sec:powercounting}. The secondary momentum dependencies of the Hedin vertices and rest functions were truncated in the spirit of the truncated unity approach introduced in~\cite{Huseman_salmhoefer_omega_flow_and_effeicient_parametrization,Lichtenstein_2017,Eckhardt_2018,multiloop_Tagliavini_2019}, and have kept only the local/$s$-wave form factors, justified by the fact that both AFM and CDW are $s$-wave orders. We have verified that the introduction of additional form factors does not change the results. The parquet equations were then solved using the strategy outlined in Fig.~\ref{fig:solution_strategy} within the PA.

\begin{figure}[h!]
    \centering
    \includegraphics[width=1.0\linewidth]{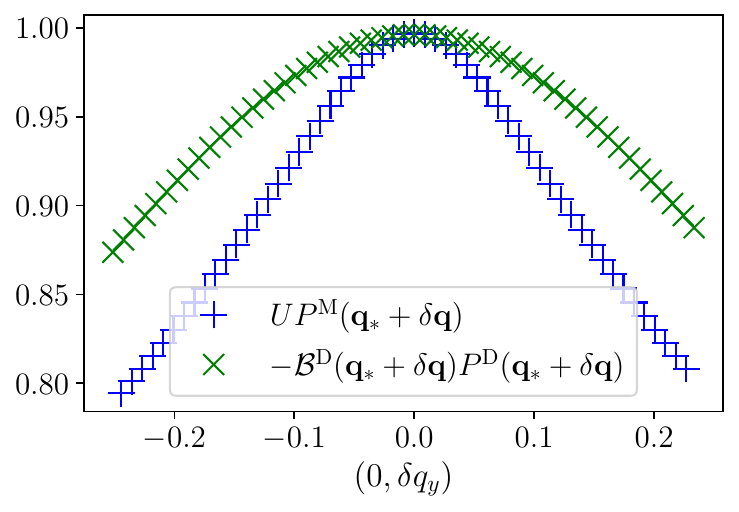}
    \caption{For the 2D Hubbard model, we plot the magnetic (density) polarizations times the bare interaction plotted in the vicinity of the ordering vector close to an AFM (CDW) instability in the PA with the self-energy neglected. The parameters are specified in the main text. }
    \label{fig:polarization_numerics}
\end{figure}

\begin{figure}[h!]
    \centering
    \includegraphics[width=1.0\linewidth]{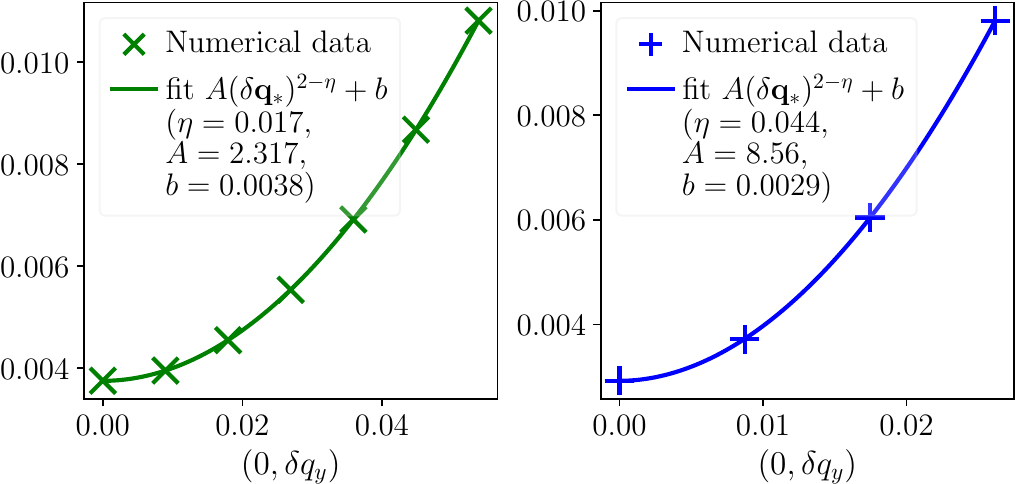}
    \caption{Fit of the critical exponent from data shown in Fig.~\ref{fig:polarization_numerics} for the anomalous dimension. We find the critical exponents $\eta \approx 0.044$ for an AFM and $\eta \approx 0.017$. }
    \label{fig:eta_fit_bruteforce}
\end{figure}

The critical exponent can be accessed from the behavior of the bosonic propagator near the ordering vector $\bfq = (\pi, \pi)$. For AFM, we expect
\begin{align}
 \expval{\varphi^{\M}(q) \varphi^{\M}(-q)} = w^{\M}(i\Omega= 0, q = q_* + \delta q) \nonumber \\ = \frac{U}{1 - P^{\M}(i\Omega = 0, \bfq_* + \delta \bfq) U} \sim \frac{1}{\abs{\delta \bfq}^{2-\eta}}. \label{eq:powerlaw_AFM}
\end{align}
Near criticality, $P^{\M}(i\Omega =0, \bfq_*)U \rightarrow 1$ leading $w^{\M}$ to display the power law behavior. We can therefore obtain $\eta$ by fitting 
\begin{align}
1 - P^{\M}(\bfq_* + \delta \bfq) U \sim  A \abs{\delta \bfq}^{2-\eta} + b \label{eq:powerlaw_inv_AFM}
\end{align}
near a critical point, where $b$ is a measure of closeness to the critical point and is zero exactly at the critical point. This is needed since in the numerics, we cannot be exactly at criticality. In order to resolve deviations from a meanfield-like Ornstein-Zernicke form, our momentum resolution $\delta \bfq$ must be such that 
\begin{align}
    \Delta \bfq < 1/\sqrt{w^\M(0, \bfq_*)} \approx \sqrt{b}. \label{eq:mom_resolution_condition}
\end{align} 
 A simple way to obtain a CDW in the repulsive  Hubbard Model on the square lattice, one needs additionally a nearest-neighbor interaction $U'$. A numerically and structurally advantageous generalization of the SBE for extended interactions was introduced in Ref.~\cite{aleryani2024screeningeffectiverpalikecharge,aleryani2025intertwinedfluctuationsisotopeeffects}. The gist of it is that one splits the bare interaction in each channel $V_0^\X = \mathcal{B}^\X + \mathcal{F}^\X$ and replaces the notion of full bare interaction-reducibility with a channel-dependent $\mathcal{B}^\X$-reducibility, where $\mathcal{B}^\X$ is the bosonic part
~\footnote{$\mathcal{B}^\X = \mathcal{B}^\X(q)$; depends only on the bosonic transfer momentum-frequency. As a result, the screened interaction also retains a pure bosonic dependence.} of the bare interaction. This ensures the screened interaction $w^\X$ remains bosonic. The density bosonic propagator then is given by 
\begin{align}
w^{\D}(q)= \nonumber \frac{\mathcal{B}^\D(\bfq)}{1 + \mathcal{B}^\D(\bfq)P^{\D}(q)}, 
\end{align}
and is related to the density susceptibility simply by
\begin{align}
\chi^\X(q) = \frac{\mathcal{B}^\X(q) - w^\X(q)}{(\mathcal{B}^\X(q))^2}. \label{eq:BF_susc_d_density_from_w_d}
\end{align}
Thus at criticality, $w^\D$ also has a power law behavior like Eq.~\eqref{eq:powerlaw_inv_AFM}, and $\eta$ can be extracted analogously.   The momentum mesh for the bosonic momenta was refined around $\bfq_*$ such that $\Delta \bfq \sim 0.01$, which we ensured is compatible with the resulting value of $b$ according to Eq.~\eqref{eq:mom_resolution_condition} in what follows. For the AFM, we have used a local Hubbard interaction $U = 2t$ and inverse temperature $\beta t = 31.5$, whereas for the CDW, we have used $U = 0.2t$ and a nearest-neighbor $U' = 0.25t$ and inverse temperature $\beta t = 8.8$. Solving the parquet equations in the PA yields results close enough to the critical point, but not too close, keeping Eq.~\eqref{eq:mom_resolution_condition} fulfilled. We show results of the product of the polarization and the bare (bosonic) interaction in Fig.~\ref{fig:polarization_numerics} over an interval in the BZ around the ordering vector.  
We observe that at the ordering vector $\delta \bfq = 0$, they are both very close to $1$, but the density term shows a wider concave curve. Subtracting 1, we can fit for the anomalous dimension as suggested by Eq.~\eqref{eq:powerlaw_inv_AFM}. For both cases, we have chosen the fitting windows such that the maximum $y$-values are similar, $\sim 0.010$. The results are shown in Fig.~\ref{fig:eta_fit_bruteforce}. 
We find that for AFM $\eta_{\text{PA}}(d = 2, N = 3) \approx 0.044$ with $b = 0.0029$ and $A = 8.56$ and for the CDW $\eta_{\text{PA}}(d = 2, N = 1) \approx 0.017$ with $b = 0.038$ and $A = 2.317$. Both of these results are rather close to $\eta = 0$, and suggest that as we approach infinite momentum resolution (representing the true thermodynamic limit where the BZ is continuous) and get closer to the critical point, they will indeed tend to zero. We have found that reducing the resolution has an effect on pushing them farther away from zero. These results suggest that for the CDW, $\eta_{\text{PA}}(d = 2, N = 1) \neq 0.453$ but is instead $0$, and it seems like the transition would be forbidden by the HMW theorem when the self-energy is included by analogous arguments to those made in Section~\ref{sec:mermin_wagner_with_self_energy} (where one uses a $\D$-channel form of the SDE instead of the $\M$-channel one).

At this stage, we emphasize an important caveat for our conclusion for the CDW~\footnote{We thank the anonymous referee for pointing this out to us.}. The recipe of diagrammatic bosonization of Section~\ref{sec:exact_bosonic_theory}, on which the mapping to the SCSA of Section~\ref{sec:connection_with_scsa} relies, does not apply for the screened interactions and polarizations that result from a $\mathcal{B}+\mathcal{F}$-splitting procedure outlined above. The obstruction arises from residual diagrams that involve $\mathcal{F}^r$. If instead one refrains from the $\mathcal{B}+\mathcal{F}$-splitting but works with SBE quantities derived from considering reducibility with respect to the entire extended bare interaction $V_0^r$ treated as a whole diagrammatic object, identification of diagrams in terms of pure bosons is possible. However, in this formulation, complications arise there in making the connection to an $O(N)$ theory without extra assumptions, see Appendix~\ref{app:extended_interactions} for further discussion.

While the $\mathcal{B}+\mathcal{F}$ approach is more efficient for numerical calculations, both approaches should yield identical susceptibilities. It follows from Eq.~\eqref{eq:BF_susc_d_density_from_w_d} that the $\mathcal{B}+\mathcal{F}$-formulated $w^\D$ should scale as $\chi^\D$ at criticality, provided $\mathcal{B}(\bfq_*) \neq 0$. Consequently, the anomalous dimension $\eta$ extracted with or without a $\mathcal{B}+\mathcal{F}$-splitting should coincide. We conclude this section by stating that drawing definitive conclusions about criticality from numerics of the type presented above is challenging, as achieving the required precision remains difficult. We leave such investigations to future work, when numerical algorithms and techniques are more advanced.

\section{Conclusion}

In this work, we have shown how individual fermionic diagrammatics in the single-boson exchange formalism can be mapped faithfully to those of a $G$-polygon bosonic theory. In particular, the screened interaction $w^r$ is identified with the bosonic propagator of this theory and the polarization $P^r$ to the bosonic self-energy. A skeleton expansion of the polarization/bosonic self-energy in the bosonic language yields a summation over bosonic interactions, identified with $G$-polygons in the fermionic theory -- see Eqs.~\eqref{eq:exact_bosonic_theory_vertices} and Figs.~\ref{fig:natural_bosonic_vertices_sbe_ph},~\ref{fig:natural_bosonic_vertices_sbe_pp}. Bearing a similarity to the trace log theory that emerges from a HST, we demonstrated how, after projecting into spin-diagonalized channels, the trace log theory can be recovered. Our diagrammatic mapping enables recasting diagrammatic many-body fermionic approximations in the language of bosons. To illustrate the utility of this, we revisited a claim of Bickers and Scalapino~\cite{parquet_critical_bickers_scalapino_1992} making a correspondence of critical PA solutions to Bray's bosonic SCSA \cite{bray_scsa_original_1974}, which was based on a speculated form of the effective bosonic theory. Appealing to power-counting arguments on our bosonic theory from a fermionic model with local interactions, we showed concretely the correspondence with the SCSA in dimensions $d \geq 2^+$, establishing $\eta_{\text{PA}} = \eta_{\text{SCSA}}(d, N)$ for a number $N$ of critical order parameter components.   

We then showed how in $d = 2$, self-consistency with the self-energy shifts the critical temperature of any transition characterized by an anomalous dimension $\eta = 0$ down to zero. This effect was argued to be numerically robust due to the presence of a negative feedback loop between the critical divergent susceptibility and the self-energy, which gets stronger as the dimension is lowered. By citing the SCSA value of $\eta(d=2^+, N=3) = 0$ for $N = 3$ AFM order parameter components, it can thus be concluded that a finite temperature second order magnetic transition is forbidden in $d = 2$ in the PA, in alignment with the HMW theorem. This conclusion should also hold for full D$\Gamma$A, which can be thought to merely shift the value of the bare interaction in a low-energy description. Moreover, it should hold for other approximations that predict a mean field $\eta = 0$ such as the FLEX or GW approximations.

For the case of a charge density wave (CDW) ordering with an $N = 1$ order parameter, results by Bray~\cite{bray_scsa_original_1974} for $\eta_{\text{SCSA}}(d = 2, N = 1)$ suggest it to be one of two values: $\eta = 0.453$ or $\eta = 0$. The former would imply that a critical solution towards CDW in $d = 2$ would be perfectly allowed in a PA solution, consistent with the HMW theorem which does not forbid ordering corresponding to an $O(1)$ order parameter. The latter value would however imply that the PA incorrectly forbids CDW.

We performed preliminary numerical calculations in $d = 2$ neglecting the self-energy, and found a value of $\eta$ closer to zero for both AFM ($N = 3$) in the pure Hubbard model. Introducing a nearest neighbor interaction $U'$ to access a CDW ($N = 1$) transition, we also found $\eta = 0$. We hope these results pave way for a future numerical investigation with more advanced numerical machinery that allow to verify this by getting closer to the critical point while maintaining a sufficiently high resolution in momentum space.

It should be stressed that since the diagrammatic bosonization scheme utilized in this work is for the SBE polarization and screened interactions, our conclusions hold only for $s$-wave ordering. Thus, the question of finding a concrete diagrammatic mapping applicable to unconventional (non-$s$-wave) order parameters, particularly those not seeded directly by the bare interaction -- such as the $d$-wave pairing instability found in the Hubbard model on the square lattice with a local/$s$-wave bare interaction $U$ -- remains open, and the conjectured connection of the PA to the SCSA in that case is yet to be clarified.

In the future, it would be interesting to study the quantum critical regimes of various approximations by retaining the frequency dependence and non-analyticities, such as those stemming from the Landau damping term in the bubble, in the expansion of the diagrammatic bosonic action. Such an extension would allow us to investigate how well various diagrammatic methods capture zero-temperature transitions, dynamical scaling, and non-Fermi-liquid behavior in strongly correlated materials.

Furthermore, on the numerical front of solving the parquet equations, promising future directions include advanced treatments of frequency dependencies, such as the intermediate representation (IR)~\cite{shinaoka2017compressing,chikano2018performance} and the discrete Lehmann representation (DLR)~\cite{kaye2022discrete,kaye2022libdlr}, as well as approaches that combine compression for both frequencies and momenta, such as the recently introduced Quantic Tensor Trains (QTT)~\cite{Shinaoka2023} and Tensor Cross Interpolation (TCI)~\cite{Ritter_2024,Rohshap2024}. These methods should open the door for more accurate numerical verification of critical exponents.

\section*{Acknowledgment}
The author thanks K.~Fraboulet, M.~Gievers, A.~Katanin, A.~Kauch, M.~Krämer, M.~Patricolo, L.~Philoxene, and J.~Profe for comments and feedback on the manuscript, and  L.~Del Re, F.~Domizio, H.~Eßl, I.~Eremin, F.~Green, S.~Heinzelmann, F.~Krien, E.~Moghadas, N.~Ritz, G.~Rohringer, A.~Toschi, and D.~Vilardi for useful discussions. I thank S.~Andergassen and M.~Scherer for their supervision and collaboration on other projects and for providing me with the environment to conduct this research.

We thank the anonymous referee for helpful comments that improved this work and for a careful reading of the manuscript.

Feynman diagrams used in this work were produced with the publicly available JaxoDraw program~\cite{jaxodraw_Binosi_2009}.

\section*{Data Availability}
Numerical data generated for this study are available from the corresponding author upon reasonable request.

\bibliography{refs}

\appendix

\onecolumngrid  

\section{Illustration of Diagrammatic Bosonisation}
\label{app:example}
We here provide two additional examples illustrating how polarization diagrams $P^\ph$ (respectively $P^\pp$) can be mapped into bosonic diagrammatics of a theory involving $\ph$ (respectively $\pp$) bosons only, namely the theory given in Eq.~\eqref{eq:exact_bosonic_theory}.

\begin{figure}[h!]
    \centering
    \begin{overpic}[width=0.9\linewidth,tics=10]{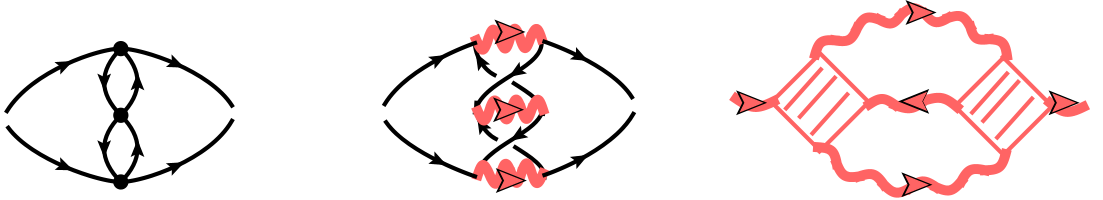}
  \put(-0.7,14){ 1)}
  \put(34,14){ 2)}
  \put(68,14){ 3)}
\end{overpic}
    \caption{Diagrammatic bosonisation example for a particle-particle polarization diagram. Resisting the temptation to identify the vertical particle-hole propagators, we identify instead the single interaction points as bare contributions to the particle-particle bosonic propagator. the result is a skeleton diagram with only even bare bosonic vertices.}
    \label{fig:example2_pp_polarization}
\end{figure}

In the first example, we consider a diagram of $P^\pp$ shown in Fig.~\ref{fig:example2_pp_polarization}. We resist the temptation to identify the particle-hole exchange subdiagram with a $\ph$ bosonic propagator, and instead, we follow the rules and only identify contributions of $w^\pp$. In this case, the three bare electron-electron interactions are the bare contributions to be identified with $w^\pp$ lines. The remaining Green's function lines form the particle-particle 4-sided polygon in Fig.~\ref{fig:natural_bosonic_vertices_sbe_pp}.

\begin{figure}[h!]
    \centering
    \begin{overpic}[width=0.9\linewidth,tics=10]{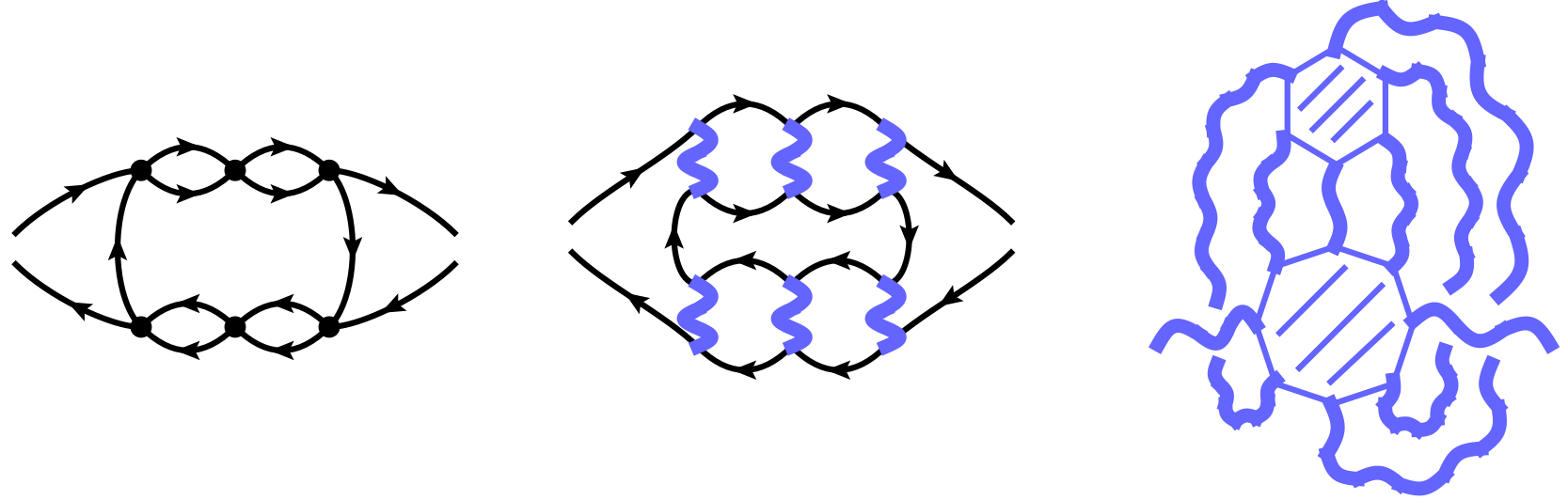}
  \put(-0.7,23){ 1)}
  \put(34,23){ 2)}
  \put(72,23){ 3)}
\end{overpic}
    \caption{Another example for bosonizing a diagram in the particle-hole polarization to produce a skeleton bosonic self-energy diagram.}
    \label{fig:example3}
\end{figure}

In the second example, we consider a contribution to $P^\ph$ shown in Fig.~\ref{fig:example3}, which feature two prominent 2-ladder particle-particle exchanges. We proceed by nonetheless identifying the bare interactions as connecting the $\pp$ bubbles by contributions of $w^\ph$. The remaining Green's function lines form closed loops that are identified with high order bosonic vertices, as illustrated in the figure. This example demonstrates how contributions of $P^\ph$ which are more naturally interpretable in terms of $\pp$ diagrams enter through diagrams with higher order vertices. It also makes explicit why approaches that truncate a single-channel HST action to low order vertices fail in situations with competing instabilities~\cite{kleinert2011hubbardstratonovichtransformationsuccessesfailure}: the feedback from a simple ladder of $k$ particle-hole bubbles onto the particle-particle channel necessarily involves diagrams containing vertices $\mathcal{V}_0^{(2n)\pp}$ with $2n \geq k$.

\begin{figure}[h!]
    \centering
    \begin{overpic}[width=0.9\linewidth,tics=10]{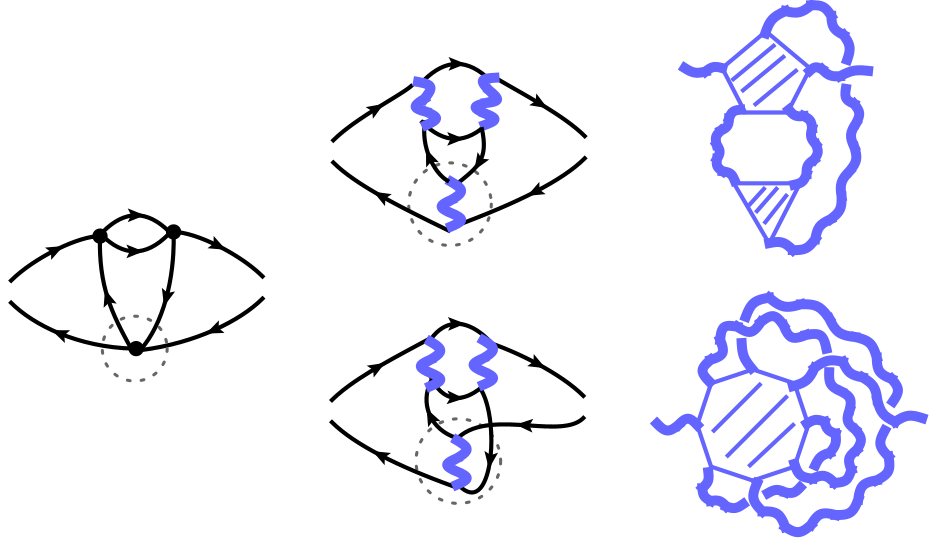}
  \put(-2,31){ 1)}
  \put(29,47){ 2)}
  \put(29,20){ 2$'$)}
  \put(65,47){ 3)}
  \put(65,20){ 3$'$)}
\end{overpic}
    \caption{In the particle-hole polarization diagrams, there are two choices of bosonic propagators to attribute to the local bare interaction. This choice reflects the multiplicity of the fermionic feynman diagrams due to the local bare interaction taking the same value for different permutations of the external legs. The different choices can lead to different bosonic diagrams.}
    \label{fig:choices}
\end{figure}

Note that a different skeleton bosonic diagram could have been chosen for example by crossing two ingoing Green's functions connecting the bosonic propagator. See Fig.~\ref{fig:choices} for an example. This is not an ambiguity, but reflects the multiplicity of the diagram and the different permutations mapping different bosonic diagrams.   Significance of multiplicity and permutations is discussed in Appendix~\ref{app:permutations_considerations} in the context of the bosonic PA.  

\begin{figure}[h!]
    \centering
    \begin{overpic}[width=0.6\linewidth,tics=10]{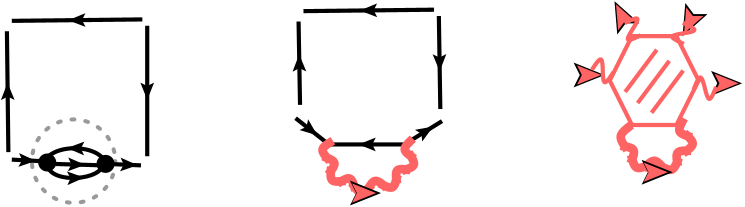}
  \put(-5,23){ 1)}
  \put(34,23){ 2)}
  \put(72,23){ 3)}
\end{overpic}
    \caption{A $GW$-type self-energy insertion, indicated with a dotted circle, in a 4-sided particle-particle $G$-polygon that would appear in the bosonization of particle-particle polarization diagram, leads to a (4+2)-sided $G$-polygon with the 2 vertexes contracted by a bosonic propagator.}
    \label{fig:self_energy_insertion_example}
\end{figure}

Finally, we discuss the role of self-energy corrections. Consider a diagram with self-energy corrections that remain after the identification of all contributions to the bosonic propagators. Such a self-energy correction would sit on a Green's function line that makes up a $G$-polygon of $n$-sides. The effect of the self-energy correction is to replace this $n$-sided $G$-polygon with an $(n + k)$ one, where the additional $k$ sides ones being mutually contracted, without affecting the other original $n - 1$ sides. More concretely, here is how we deal with bosonizing this self-energy correction. Inserting Eq.~\eqref{eq:hedin_vertex} into Eq.~\eqref{eq:SDE_SBE}, we can see find its general structure is
\begin{align}
\mathclap{\int_{k'q\sigma\sigma'}} (w^r(q)B^r_{k'k}(q))_{\sigma_1\sigma;\sigma'\sigma_2} G(q-k)_{\sigma \sigma'},
\end{align}
with $B^r_{k'k}(q) = \mathbf{1}^r$ for a $GW$-type correction, or $B^r_{k'k}(q) = \Pi^r_{\text{s-wave}, k'}(q) A^r_{k'(q-k)}(q)$ for a vertex correction. Here $A^r$ represents, as before, a $U$-irreducible diagram in the channel $r$, and can be bosonized following the instructions outlined in Section~\ref{sec:exact_bosonic_theory}. An example is shown in Fig.~\ref{fig:self_energy_insertion_example} for a self-energy correction on a particle-particle bosonic vertex which would occur in bosonizing a particle-particle polarization diagram. 

\section{Consideration of permutations of the bare bosonic vertex labels in critical PA}
\label{app:permutations_considerations}
\begin{figure}[h!]
    \centering
    \includegraphics[width=0.7\linewidth]{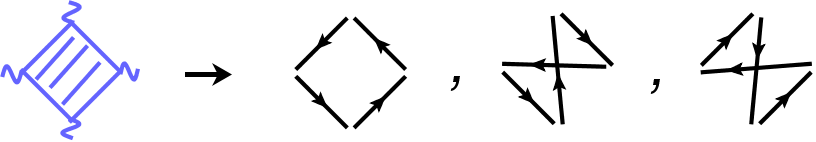}
    \caption{The labels of the bare bosonic vertex in the fermionic language can be effectively permuted in three diagrammatically distinct ways: the "square" permutation and two "hourglass" like permutations.}
    \label{fig:bosonic_bare_vertex_permutations}
\end{figure}

In this section, we investigate the role of label permutations of the bare bosonic vertex in a critical PA theory. Without loss of generality, we consider criticality in the particle-hole channel ($r_* = \ph$). An analogous analysis can be done for the $r_* = \pp$ case. Given a 4-point bosonic vertex, there are $4! = 24$ ways to connect distinct bosonic propagators to it. One may -- equivalently to permuting the connections of the propagators -- permute the labels of the corners of the vertex. This relabeling can also be thought of as changing the location of the corners of the vertex. Of the 24 ways, there are three diagrammatically distinct ways to do so corresponding to replacing the bare bosonic vertex with a "square" and two "hourglass" forms, shown in Fig.~\ref{fig:bosonic_bare_vertex_permutations}. The remaining 21 permutations can be produced by two reflections and four rotations on these three, and up to possible reversal of the fermionic arrows look the same to the aforementioned three diagrammatically. 

 When analyzing the bosonic self-energy diagrams and whether they are included in the critical PA polarization in Section~\ref{sec:connection_with_scsa}, we only considered the square permutation, and ignored the other two hourglass ones. The important point is that changing the permutations of the bare bosonic vertex labels can change the topology of the corresponding fermionic diagrams. Nonetheless, we will see here that this will not change the conclusions of the main text in a crucial way.

\begin{figure}[h!]
    \centering
    \includegraphics[width=0.8\linewidth]{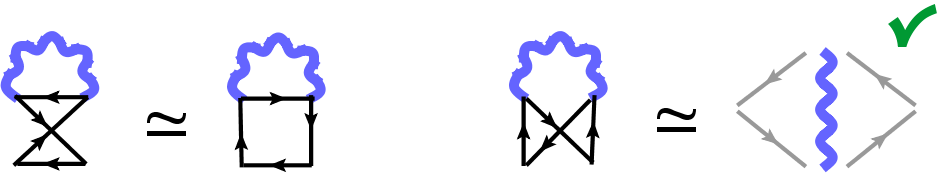}
    \caption{In the translating the bosonic tadpole diagram to the fermionic language - see also Fig.~\ref{fig:hartree-fock_boson} - one should consider additionally the two diagrammatically distinct hourglass permutations. The first of them leads to the same conclusion surrounding Fig.~\ref{fig:hartree-fock_boson}. The second however gives a genuine polarization diagram. The grey GG pairs are identified with the two bubbles that appears in the correlated part of the polarization in Eq.~\eqref{eq:polarization_layedout_correlated}. }
    \label{fig:hartreefock_permutations_variant}
\end{figure}

Let us begin with the tadpole O$(\mathcal{V}_0)$ term and insert instead of the square permutation as in Fig.~\ref{fig:hartree-fock_boson}, the two hourglass contributions. We find that one of the hourglass permutations leads to a topologically equivalent diagram to the one with the square variant. The conclusion is also the same; such a diagram is not skeleton in the fermion propagators and can be removed by a redefinition of the fermionic bare self-energy. The second hourglass permutation however does lead to a contribution of the polarization that is also a PA diagram. See Fig.~\ref{fig:hartreefock_permutations_variant}. Thus, a third of the 24 permutations of the tadpole diagrams are included. 

\begin{figure}[h!]
    \centering
    \includegraphics[width=0.8\linewidth]{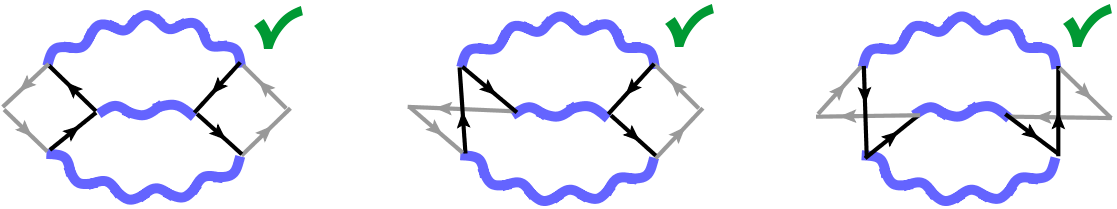}
    \caption{Considering any of the three diagramatically distinct permutations in translating the second order bosonic self-energy contribution yield in the fermionic language to a PA polarization diagram, as can be checked explicitly by cutting $GG$ lines as described in the caption of Fig.~\ref{fig:fourth_order_pa_not_pa} and the discussion around it.  }
    \label{fig:secondorder_permutations_variant}
\end{figure}

\begin{figure}[h!]
    \centering
    \includegraphics[width=0.99\linewidth]{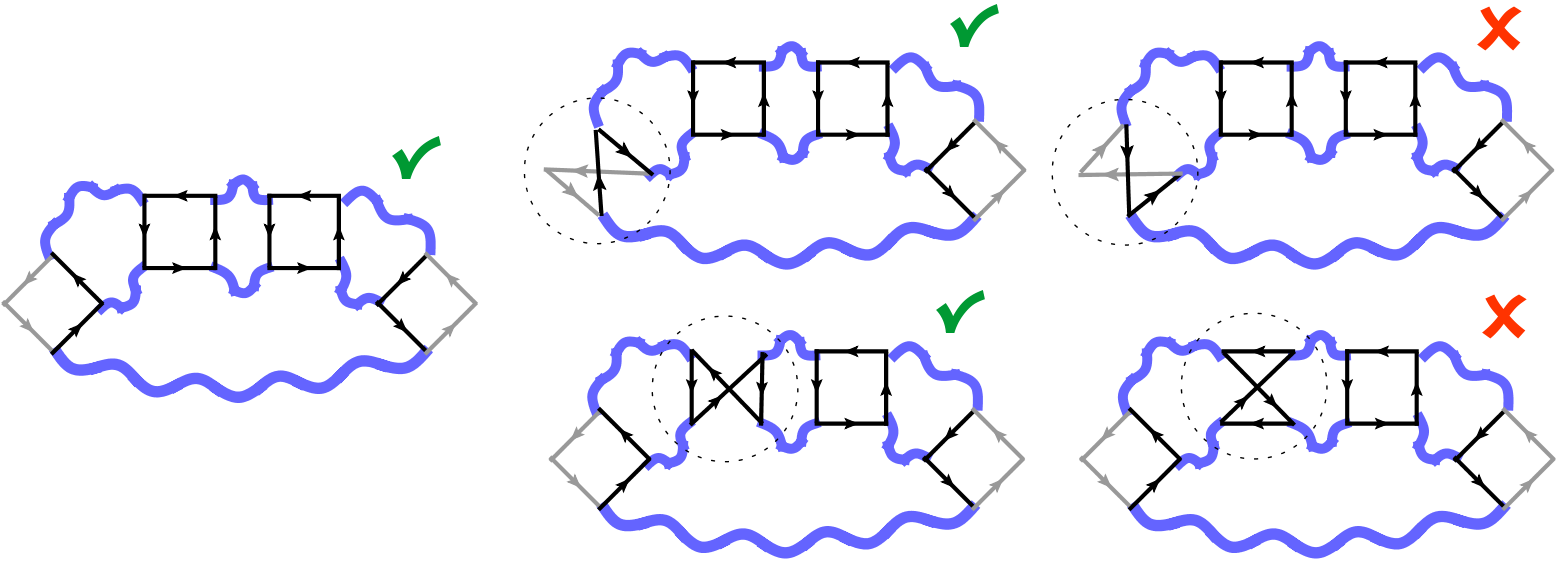}
    \caption{For the third order and higher bosonic diagrams, only the square and one of the two hourglass permutations yield a PA polarization diagrams as can be explicitly verified with the examples marked with the green check mark. Introducing an hourglass permutation along the rainbow leads to a non PA diagram as can be explicitly verified with the examples marked with the red cross. }
    \label{fig:higherorder_permutations_variant}
\end{figure}

Next, we consider the diagrams of order $\mathcal{V}^2$. We find that replacement either of the the two bosonic bare vertices with any of the three permutations still yield PA diagrams. See Fig.~\ref{fig:secondorder_permutations_variant}.
At order $\mathcal{V}_0^3$ and higher skeleton contributions, one of the two hourglass contributions lead to a non-PA contribution wherever it is placed, whereas the other still lead to a PA diagram in the fermionic language, see Fig.~\ref{fig:higherorder_permutations_variant} for examples. All non-rainbow diagrams that appear at order $\mathcal{V}_0^4$ remain non-PA regardless of which permutations are used. 

\begin{figure}[h!]
    \centering
    \includegraphics[width=0.9\linewidth]{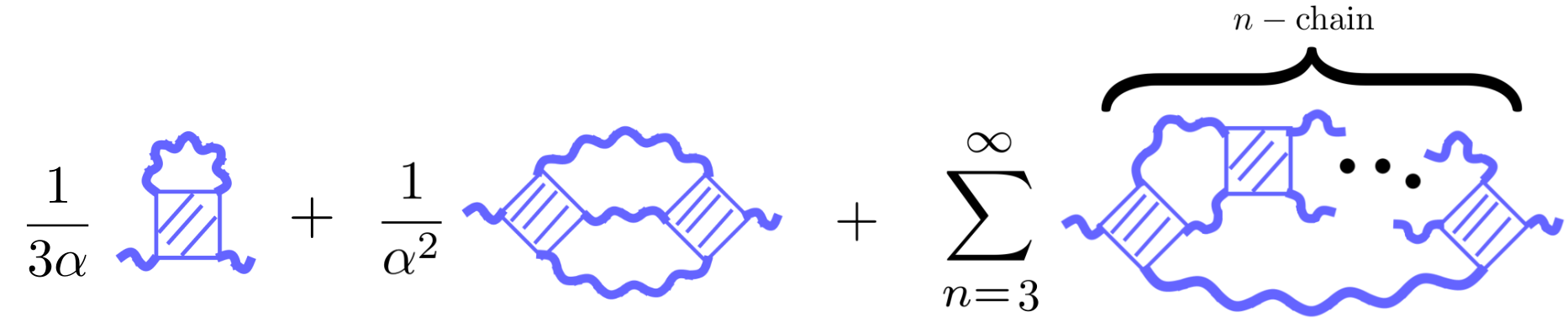}
    \caption{Full consideration of the permutations of the connections to the bare bosonic vertices show that the skeleton critical PA polarization in the bosonic language is a sum of rainbow diagrams, with a difference from Fig.~\ref{fig:rainbow_summation} that the first- and second- order term have relative prefactors. Such a difference cannot change the universal behavior, and the critical PA is thus still expected to share the universal critical exponents of the SCSA. }
    \label{fig:modified_scsa}
\end{figure}

This means that the bosonic viewpoint, PA diagrams beyond the second order contain only two thirds of the 24 permutations for the bare vertices. As far as critical aspects of the PA are considered, the bare point vertex can be treated as a constant (follows from power-counting in Section.~\ref{sec:powercounting} ), which means we need only scale the bare vertices appearing in the rainbow skeleton diagrams at order $\mathcal{V}^3$ and above by a compensation factor 
\begin{align}
    \mathcal{V} \rightarrow \alpha \mathcal{V} \qq{with } \alpha = 2/3.
\end{align} 
Equivalently, we alternatively scale the first- and second- order diagrams by a factor of $\alpha^{-1}$ for every bare vertex that appears in them. Thus, in the bosonic viewpoint, the bosonic self-energy of the critical PA solution will be given by the summation shown in Fig.~\ref{fig:modified_scsa}. Note that the first- order diagram comes with an additional prefactor of $1/3$ as a consequence of the previous discussion. The result differs from the critical SCSA solution for the self-energy by the factors that sits next to the first- and second- order terms. Since by themselves, both of these term are analytic, this cannot change the universal behaviour of the approximation. In particular, we expect the critical exponents and scaling laws to not change with respect to the SCSA. 

\section{Comments on diagrammatic bosonization for models with extended interactions}
\label{app:extended_interactions}
We comment briefly on what happens if the model considered features an extended interaction $V_0$ rather than the local one in Eq.~\eqref{eq:bare_vertex_spin_invt} considered thus far. For a density-density interaction, it would take the form
\begin{align}
V_0 = \delta_{\bfk_1 + \bfk_3, \bfk_2 +\bfk4}\left(v_0(\bfk_2 - \bfk_1)\delta_{\sigma_1,\sigma_2} \delta_{\sigma_3,\sigma_4} - v_0(\bfk_4 - \bfk_1)\delta_{\sigma_1, \sigma_4} \delta_{\sigma_3, \sigma_2}\right).\label{eq:bare_interaction_extended}
\end{align}
In general, such an extended interaction will maintain momentum dependence in the three channels $U \rightarrow V_{0;\bfk \bfk'}^r(\bfq)$
and running the SBE recipe with respect to $V_0^r$-reducibility will yield screened interactions that will also depend on secondary momenta $\bfk$ and $\bfk'$
\begin{align}
w^r = w^r_{\bfk \bfk'}(q).
\end{align}
This screened interaction is not purely bosonic. The situation simplifies slightly if the bare interaction involves only a finite set of neighbors. For example, in the extended Hubbard model which includes interaction to nearest neighbors, one has
\begin{align}
v_0(\bfq) \equiv U + 2U'\sum_{i=1}^d\cos(\bfq_i),\label{eq:extended_hubbard_cube_bare_interaction}
\end{align}
with $d = 2$ for the square lattice (for a cubic lattice, $d = 3$). Then the bare interaction has a finite form factor decomposition
\begin{align}
V^r_{0;\bfk\bfk'}(\bfq) \equiv \int_{lm} V^r_{0;lm}(\bfq) f^*_l(\bfk') f_{m}(\bfk),
\end{align}
where here $\{f_m(\bfk)\}$ are a minimal \emph{finite} set of form factors. In the case of the interaction specified by Eq.~\ref{eq:extended_hubbard_cube_bare_interaction}, one can express the bare interaction exactly with five form factors for the onsite and first shell ($s$-wave with $f_m(\bfk) \equiv 1$, $d$-wave with $f_m(\bfk) \equiv \cos \bfk_x - \cos \bfk_y$, $p_x$-wave with $f_m(\bfk) \equiv \sin \bfk_x$, $p_y$-wave with $f_m(\bfk) \equiv \sin \bfk_y$ and extended-$s$-wave with $f_m(\bfk) \equiv \cos \bfk_x + \cos \bfk_y$). Consequently, if we write a screened interaction in form factor space, it is a $5\times5$ matrix $w^r_{lm}(q)$ playing the role of a matrix-valued bosonic propagator. In particular, each of the five diagonal entries $w^r_{mm}(q)$ plays the role of the propagator of a type of a boson, and the corresponding order parameter field $\varphi^{r}_m$ transforms according to the symmetry of the corresponding form factor $f_m$. The polarization $P^r(q)$ is then also a $5\times5$ matrix, playing the role of a matrix-valued self-energy. The bosonic vertices $\mathcal{V}_0^{(n)r}$ acquire $n$ additional indices $m_j$ indicating one of the five types of incoming or outgoing bosons at each corner. In their graphical representation similar to Figs.~\ref{fig:natural_bosonic_vertices_sbe_pp} and \ref{fig:natural_bosonic_vertices_sbe_ph}, each vertex expands into a family of vertices encompassing all corner types beyond the simple ``$s$-wave''. As a result, distinct vertices arise in which different corners can couple to different bosonic species. The analytical expression for these vertices will involve a form factor insertion next to each propagator in Eq.~\eqref{eq:exact_bosonic_theory_vertices}. In cases of criticality in a given channel $r_*$, one of the eigenvalues (or several degenerate ones) of the matrix $w^{r_*}_{lm}$ at some ordering vector will diverge. In this case, since the different components of the field $\varphi^r_m$ are correlated ($\expval{\varphi^r_m \varphi^r_l} \equiv w^r_{ml}\neq 0$ for $m \neq l$ in general), it is less straightforward to connect the theory to an $O(N)$ model, -- and thus the PA to the SCSA, -- at criticality, without extra assumptions.

\section{Alternative line of argument that self-energy convergence in $d = 2$ is inconsistent with finite temperature criticality with $\eta = 0$}
\label{app:alternative_se_proof}
In the main text, we gave an argument using the triangular form of the SDE in terms of the SBE quantities that in $d = 2$, full consistency in self-energy demands that finite temperature criticality with $\eta = 0$ are not possible. In the argument, we used that $\lambda^\X \approx \text{const} \neq 0$. Here, we give a different more elaborate argumentation based on the traditional parquet equations that does not rely on that. 

In physical channels/spin-diagonalized representation, the generalized susceptibility fullfills the following Bethe-Salpeter equation~\cite{Rohringerthesis} 
\begin{align}
\chi_{k k'}^\X(q) &= \Pi^\X_{k k'}(q) + \left[\chi^\X(q) \circ I^\X(q) \circ \Pi^\X(q)\right]_{kk'}, \label{eq:generalised_susc}
\end{align}
with $I^\X$ the $GG$-irreducible vertex in the channel $\X$. The composition symbol $\circ$ represents matrix multiplication in secondary momenta $k$ and $k'$. Eq.~\eqref{eq:generalised_susc} is formally solved by
\begin{align}
\chi^\X_{kk'}(q) = \left[ \Pi^\X(q)\circ \left(\mathbf{1} -  I^\X(q)\circ \Pi^\X(q)\right)^{-1} \right]_{kk'}, \label{eq:bse_susc_soln}
\end{align}
where the inverse is the matrix inversion, and the identity matrix in secondary momenta has components $[\mathbf{1}]_{kk'} = \delta_{kk'}$.
Since $\Pi^\X$ is finite and non-zero, criticality is signaled with the inverted matrix showing in the expression developing a zero eigenvalue. The so-called Bethe-Salpeter kernel $I^\X(q)\circ \Pi^\X(q)$ can be diagonalized
\begin{align}
\left[I^\M(q) \circ \Pi^\M(q)\right]_{kk'} = \int_{\gamma}  \alpha_\gamma(q) R_k^\gamma(q) L^{\gamma}_{k'}(q)^*,
\end{align}
where, without loss of generality, we have specialized to the magnetic channel for illustration in what follows. Here, $L^\gamma$ and $R^\gamma$ are ($q$-dependent) left and right eigenvectors labeled by $\gamma$, and $\alpha^\gamma$ are the ($q$-dependent) eigenvalues. The two matrices are biorthogonal; i.e.
\begin{align}
\int_k L_k^\gamma(q)^* R_{k}^{\gamma'}(q) =  \delta_{\gamma\gamma'}.
\end{align}
For an object $A_{kk'}(q)$, we introduce the notation
\begin{align}
A_{\gamma\gamma'}(q) = \left[L(q)^\dagger\circ A(q) \circ R(q)\right]_{\gamma\gamma'} = \int_{kk'}L_{k}^\gamma(q)^* A_{kk'}(q) R_{k'}^{\gamma'}(q). 
\end{align}
That $L$ and $R$ diagonalize $I^\X\circ \Pi^\X$ mean that $\left[ I^\X\circ\Pi^\X \right]_{\gamma\gamma'} = \alpha_{\gamma} \delta_{\gamma\gamma'}$.
Multiplying both sides of Eq.~\eqref{eq:bse_susc_soln} from the left by $L^\dagger$ and from the right by $R$, we thus find
\begin{align}
\chi^\M_{\gamma\gamma'}(q) = \frac{\Pi_{\gamma\gamma'}(q)}{1- \alpha^\gamma(q)},
\end{align}
and we see that at a critical point corresponding to an eigenvalue $\alpha_{\gamma_*}$ with ordering vector $q_*$, the eigenvalue $\alpha_{\gamma_*}(q_*) \rightarrow 1$. This equation can be thought of as a generalization of the expression of the s-wave magnetic susceptibility
\begin{align}
\chi^\M(q) = \frac{U}{1- UP^\M(q)},
\end{align}
where $P^\M$ is the SBE polarization. In this respect the SBE parametrization avoids the need of diagonalizing the Bethe-Salpeter kernel, but the connection to the screened interactions and polarizations that allows it is limited to s-wave susceptibilities. We also have the scaling behaviour at criticality in the vicinity of the ordering vector $\delta \bfq$
\begin{align}
1 - \alpha(q_* + \delta q) \sim \abs{\bfq}^{2-\eta}, 
\end{align}
in analogy to Eq.~\eqref{eq:powerlaw_inv_AFM}. The Bethe-Salpeter kernel also shows up in the Bethe-Salpeter equations for the vertex within the traditional Parquet equations
\begin{align}
    V^\X(q) &= I^\X(q) + I^\X(q) \circ \Pi^\X(q) \circ V^\X(q) \nonumber \\
    &= (\mathbf{1} - I^\X(q) \circ \Pi^\X(q))^{-1} \circ I^\X(q). \label{eq:BSE_vertex}
\end{align}
For the self-energy, the traditional form of the SDE can be written in terms of a single channel notation as follows 
\begin{align}
\Sigma(k) - \Sigma_{\text{HF}} &= -\int_{k'' q}V^\M_{kk''}(q) G(k''+q) G(k'')G(k + q)U \nonumber \\
&= \int_{k''q\tilde{k}}V^\M_{kk''}(q) \Pi^\M_{k''\tilde{k}}(q)\delta_{\tilde{k}k''} G(k + q)U \nonumber \\
&=\int_{k''q}\left[ (1 - I^\M\circ \Pi^\M)^{-1}\circ I^\M \circ \Pi^\M \right]_{kk''}  G(k + q)U \nonumber \\
&=\int_{k''q}\int_{\gamma} \frac{\alpha_\gamma(q)}{1 - \alpha_\gamma(q)}  R^\gamma_k(q)  L_{k''}^{\gamma}(q)^*  G(k + q)U\nonumber \\
&=U\int_{q}\int_{\gamma} \frac{\alpha_\gamma(q)}{1 - \alpha_\gamma(q)}  \left(G(k + q)\int_{k''} R^\gamma_k(q) L_{k''}^{\gamma}(q)^*\right),
\end{align}
where we have used Eqs.~\eqref{eq:BSE_vertex} to bring out the Bethe-Salpeter kernel and expressed it in terms of its eigenvalue decomposition. Assuming criticality in a specific eigenvalue label $\gamma_*$, we write
\begin{align}
\Sigma(k) &\approx \Sigma_{\text{regular}}(k) + U\int_{q} \frac{\alpha_{\gamma_*}(q)}{1 - \alpha_{\gamma_*}(q)}  \left(G(k + q) R^{\gamma_*}_k(q) \int_{k''}L_{k''}^{\gamma_*}(q)^*\right)\\
& \approx \Sigma_{\text{regular}}(k) + U\int_{\abs{\delta \bfq} < \epsilon} \frac{1}{\abs{\delta \bfq}^{2-\eta}}  \left(G(i\nu, \bfk + \bfq_*) R^{\gamma_*}_k(\bfq_*) \int_{k''}L_{k''}^{\gamma_*}(\bfq_*)^*\right)
\end{align}
where we have neglected the contributions of the noncritical eigenvalues, substituted the scaling form and split the integral similar to Eq.~\eqref{eq:split_sig_integrand}. Next, we claim that if the instability is of the s-wave type, then the momentum $\bfk$ dependence of $L^{\gamma_*}_k(q_*)$ and $R^{\gamma_*}_k(q_*)$ is weak. To see this, note that the s-wave susceptibility is simply the integral over the $k$ and $k'$ dependencies of the generalized susceptibility
\begin{align}
\chi^\M(q) = \int_{kk'} \chi^\M_{kk'}(q),
\end{align}
For an s-wave instability, we expect the critical eigenvalues to correspond to the s-wave susceptibility; $\chi^\M_{\gamma_*\gamma_*}(q) \approx \chi^\M(q)$, which suggest that $L^{\gamma_*}(q_*)$ and $R^{\gamma_*}(q_*)$ are close to being constants as functions of $k$ and $k'$, and we reduce to the expression in Eq.~\eqref{eq:sig_critical_expr}.

\end{document}